%% file: paper_v3.tex
\documentclass[submission,letterpaper,Phys]{SciPost}
\pdfoutput=1
\usepackage{amsmath,amssymb,mathtools,xspace}
\usepackage{booktabs,multirow,graphicx,tabularx,slashed}
\usepackage{hyperref}
\usepackage{color,xcolor}
\usepackage[normalem]{ulem}
\usepackage{enumitem}
\usepackage{feynmp}
\usepackage{braket,stackrel}
\usepackage{tikz}
\usetikzlibrary{arrows}
\usetikzlibrary{shapes.geometric, arrows}

\makeatletter
\@ifundefined{pdfoutput}{}{\DeclareGraphicsRule{*}{mps}{*}{}}
\makeatother

\makeatletter
\DeclareRobustCommand*{\bfseries}{%
   \not@math@alphabet\bfseries\mathbf
   \fontseries\bfdefault\selectfont
   \boldmath
}
\makeatother

\input{incl_declare}

\begin{document}
\begin{fmffile}{feynman}

\begin{center}{\Large \textbf{
Invertible Networks or Partons to Detector and Back Again
}}\end{center}

\begin{center}
Marco Bellagente\textsuperscript{1},
Anja Butter\textsuperscript{1},
Gregor Kasieczka\textsuperscript{3},
Tilman Plehn\textsuperscript{1}, 
Armand Rousselot\textsuperscript{1,2},
Ramon Winterhalder\textsuperscript{1},
Lynton Ardizzone\textsuperscript{2}, and
Ullrich K\"othe\textsuperscript{2}
\end{center}

\begin{center}
{\bf 1} Institut f\"ur Theoretische Physik, Universit\"at Heidelberg, Germany\\
{\bf 2} Heidelberg Collaboratory for Image Processing, Universit\"at Heidelberg, Germany\\
{\bf 3} Institut f\"ur Experimentalphysik, Universit\"at Hamburg, Germany
butter@thphys.uni-heidelberg.de
\end{center}

\begin{center}
\today
\end{center}

% For convenience during refereeing: line numbers
%\linenumbers

\tikzstyle{int}=[thick,draw, minimum size=2em]
%\tikzstyle{init} = [pin edge={to-,thin,black}]

\section*{Abstract}
{\bf For simulations where the forward and the inverse directions have
  a physics meaning, invertible neural networks are especially useful.
  A conditional INN can invert a detector simulation in terms of
  high-level observables, specifically for ZW production at the
  LHC. It allows for a per-event statistical interpretation. Next, we
  allow for a variable number of QCD jets. We unfold detector effects
  and QCD radiation to a pre-defined hard process, again with a
  per-event probabilistic interpretation over parton-level phase
  space.}

\vspace{10pt}
\noindent\rule{\textwidth}{1pt}
\tableofcontents\thispagestyle{fancy}
\noindent\rule{\textwidth}{1pt}
%\vspace{10pt}

\newpage
%%%%%%%%%%%%%%%%%%%%%%%%%%%%%%%%%%%%%%%%%%%%%%%%%%%%%%%%%%%%%%%%%%%%%%%%
\section{Introduction}
\label{sec:intro}

The unique feature of LHC physics from a data science perspective is
the comparison of vast amounts of data with predictions based on first
principles. This modular prediction starts with the Lagrangian
describing the hard scattering, then adds perturbative QCD providing
precision predictions, resummed QCD describing parton showers and
fragmentation, hadronization, and finally a full detector
simulation~\cite{black_book}. In this so-defined forward direction all
simulation modules are based on Monte Carlo techniques, and in the
ideal world we would just compare measured and simulated events and
draw conclusions about the hard process. This hard process is where we
expect to learn about new aspects of fundamental physics, for instance
dark matter, extended gauge groups, or additional Higgs bosons.

Because our simulation chain works only in one direction, the typical
LHC analysis starts with a new, theory-inspired hypothesis encoded in
a Lagrangian as new particles and couplings. For every point in the
new physics parameter space we simulate events, compare them to the
measured data using likelihood methods, and discard the new physics
hypothesis. This approach is inefficient for a variety of reasons:
\begin{enumerate}
  \item The best way to compare two hypotheses is the log-likelihood
    ratio based on new physics and Standard Model predictions for the
    hard process. Using this ratio in the analysis is the idea behind
    the matrix element
    method~\cite{Kondo:1988yd,Martini:2015fsa,Gritsan:2016hjl,Martini:2017ydu,Kraus:2019qoq,Prestel:2019neg},
    but usually this information is not
    available~\cite{Cranmer:2019eaq}.
\item New physics hypotheses have free model parameters like masses or
  couplings, even if an analysis happens to be independent of them.
  If the predicted event rates follow a simple scaling, like for a
  truncated effective theory, this is simple, but usually we need to
  simulate events for each point in model space.
\item There is a limit in electroweak or QCD precision to which we can
  reasonably include predictions in our standard simulation
  tools. Beyond this limit we can, for instance, only compute a
  limited set of kinematic distributions, which excludes these
  precision prediction from standard analyses.
\item Without a major effort it is impossible for model builders to
  derive competitive limits on a new model by recasting an existing
  analysis.
\end{enumerate}
All these shortcomings point into the same direction: we need to
invert the simulation chain, apply this inversion to the measured
data, and compare hypotheses at the level of the hard scattering. For
hadronization and fragmentation an approximate inversion is standard
in that we always apply jet algorithms to extract simple parton
properties from the complex QCD jets.  For the detector simulation
either at the level of particles or at the level of jets this problem
is usually referred to as detector unfolding. For instance in top
physics we also unfold kinematic information to the level of the
decaying top quarks, assuming that the top decays are correctly
described by the standard
model~\cite{Khachatryan:2015oqa,Aad:2015mbv}. Going beyond detector
effects we know what for many analyses QCD jet radiation adds little
to our new physics search. This is certainly true whenever soft and
collinear radiation can be simulated by spin-averaged parton showers
depending only logaritmically on the global energy scale of the hard
process. In that case we should also be able to also unfold QCD jet
radiation as the last simulation step. This is the final goal of our
paper.

Technically, we propose to use invertible networks
(INNs)~\cite{inn1,coupling2,glow} to invert part of the LHC simulation
chain.  This application builds on a long list of one-directional
applications of generative or similar networks to LHC simulations,
including phase space integration~\cite{maxim,bendavid},
amplitudes~\cite{Bishara:2019iwh,Badger:2020uow}, event
generation~\cite{dutch,gan_datasets,DijetGAN2,gan_phasespace,Alanazi:2020klf},
event subtraction~\cite{subgan}, detector
simulations~\cite{calogan1,calogan2,fast_accurate,aachen_wgan1,aachen_wgan2,ATLASShowerGAN,ATLASsimGAN,Belayneh:2019vyx,Buhmann:2020pmy},
parton showers~\cite{shower,locationGAN,monkshower,juniprshower}, or
searches for physics beyond the Standard Model~\cite{bsm_gan}.  INNs
are an alternative class of generative networks, based on normalizing
flows~\cite{nflow1,papamakarios2019normalizing,nflow_review,mller2018neural}.
In particle physics such normalizing flow networks have proven useful
for instance in phase space generation~\cite{Bothmann:2020ywa},
linking integration with generation~\cite{Gao:2020vdv,Gao:2020zvv}, or
anomaly detection~\cite{Nachman:2020lpy}.

Our INN study on unfolding detector-level
events~\cite{Andreassen:2019cjw} to the hard scattering builds on
similar attempts with a standard GAN~\cite{Datta:2018mwd} and a fully
conditional GAN analysis~\cite{fcgan}.  In Sec.~\ref{sec:inn} we show
how the bijective structure of the INN makes their training especially
stable.  If we add sufficiently many random numbers to the INN we can
start generating probability distributions in the parton-level phase
space.  The conditional INN (cINN)~\cite{cinn,cinn2} adds even more
sampling elements to the generation of unfolded configurations.  For
arbitrary kinematic distributions we can test the calibration of this
generative network output using truth information and find that unlike
GANs the cINN lives up to its generative promise: for a single
detector-level event the cINN generates probability distributions in
the multi-dimensional parton-level phase space.

Next, we show in Sec.~\ref{sec:jets} how the inversion can link two
phase spaces with different dimension. This allows us to unfold based
on a model with a variable number of final state particles at the
detector level and is crucial to include higher-order perturbative
corrections. We show how the cINN can account for jet radiation and
unfolds it together with the detector effects.  In other words, the
network distinguishes between jets from the hard process and jets from
QCD radiation and it also unfolds the kinematic modifications from
initial state radiation, to provide probability distributions in the
parton-level phase space of a hard process.

We note that our examples only cover analyses where subjet
information factorizes from the hard process, for instance in terms of
(mis-)tagging efficiencies. For analyses going beyond this level, like
searches for long-lived particles, we need to skip the jet algorithm
stage and instead include the full calorimeter and tracking
information. In principle and assuming the availability of a proper
detector simulations our ideas might still work for these
applications, but for the time being we ignore these complications.

%%%%%%%%%%%%%%%%%%%%%%%%%%%%%%%%%%%%%%%%%%%%%%%%%%%%%%%%%%%%%%%%%%%%%%%%
\section{Unfolding basics}
\label{sec:basics}

Unfolding particle physics events is a classic example for an inverse
problem~\cite{Gagunashvili:2010zw, Spano:2013nca, Glazov:2017vni}. In the limit where detector effects can
be described by Gaussian noise, it is similar to unblurring
images. However, actual detector effects depend on the individual
objects, the global energy deposition per event, and the proximity of
objects, which means they are much more complicated than Gaussian noise. The
situation gets more complicated when we add effects like QCD jet
radiation, where the radiation pattern depends for instance on the
quantum numbers of the incoming partons and on the energy scale of the
hard process.

What we do know is that we can describe the measurement of phase space detector-level distributions $\d \sigma/\d x_d$ as a random process, just as the
detector effects or jet radiation can be simulated by a set of random
numbers describing a Markov process. This means that also the
inversion or extraction of the parton-level distribution $\d \sigma/\d x_p$ is a statistical problem.

%%%%%%%%%%%%%%%%%%%%%%%%%%%%%%%%%%%%%%%%%%%%%%%%%%%%%%%%%%%%%%%%%%%%%%%%
\subsection{Binned toy model and locality}
\label{sec:basics_binned}

As a one-dimensional toy example we can look at a binned
(parton-level) distribution $\sigma^{(p)}_j$ which gets transformed
into another binned (detector-level) distribution $\sigma^{(d)}_j$ by
the kernel or response function $g_{ij}$,
\begin{align}
  \sigma^{(d)}_i = \sum_{j=1}^N g_{ij} \sigma^{(p)}_j \; .
\end{align}
We can postulate the existence of an inversion with the kernel
$\bar{g}$ through the relation
\begin{align}
  \sigma^{(p)}_k
  = \sum_{i=1}^N \bar{g}_{ki} \sigma^{(d)}_i 
%  = \sum_i \bar{g}_{ki} \sum_j g_{ij} \sigma^{(p)}_j \notag \\
  = \sum_{j=1}^N \left( \sum_{i=1}^N \bar{g}_{ki} g_{ij} \right) \sigma^{(p)}_j 
  \quad \text{with} \quad
  \sum_{i=1}^N \bar{g}_{ki} g_{ij} = \delta_{kj} \; .
\end{align}
If we assume that we know the $N^2$ entries of the kernel $g$, this
form gives us the $N^2$ conditions to compute its inverse $\bar{g}$.
We illustrate this one-dimensional binned case with a semi-realistic
smearing matrix
\begin{align}
  g
  &=
  \begin{pmatrix}
  1 - x & x & 0 \\ x & 1-2x & x \\ 0 & x & 1-x
  \end{pmatrix} \; .
\label{eq:gtoy}
\end{align}
We illustrate the smearing pattern with two input vectors, keeping in
mind that in an unfolding problem we typically only have one kinematic
distribution to determine the inverse matrix $\bar{g}$,
\begin{alignat}{7}
  \sigma^{(p)} &= n \begin{pmatrix} 1 \\ 1 \\ 1 \end{pmatrix}
  &\quad &\Rightarrow \quad
  \sigma^{(d)} = \sigma^{(p)}\; , \notag \\
  \sigma^{(p)} &= \begin{pmatrix} 1 \\ n \\ 0 \end{pmatrix}
  &\quad &\Rightarrow \quad
  \sigma^{(d)} %= \begin{pmatrix} 1+(n-1)x \\ n-2(n-1)x \\ 1+(n-1)x \end{pmatrix}
  = \sigma^{(p)} + x \begin{pmatrix} n-1 \\ -2n + 1 \\ n \end{pmatrix} \; .
\end{alignat}
The first example shows how for a symmetric smearing matrix a flat
distribution removes all information about the detector effects. This
implies that we might end up with a choice of reference
process and phase space such that we cannot extract the
detector effects from the available data. The second example
illustrates that for bin migration from a dominant peak the
information from the original $\sigma^{(p)}$ gets overwhelmed
easily. We can also compute the inverse of the smearing matrix in
Eq.\eqref{eq:gtoy} and find
\begin{align}
  \bar{g}
%  &= \frac{1}{(3x -1)(x-1)} 
%  \begin{pmatrix}
%  x^2 -3x +1 & x(x-1) & x^2 \\ x(x-1) & (x-1)^2 & x(x-1) \\ x^2 & x(x-1) & x^2-3x-1
%  \end{pmatrix}
%  &= \frac{1}{(1-3x)(1-x)} 
%  \begin{pmatrix}
%  1-3x+x^2 & -x+x^2 & x^2 \\ -x+x^2 & (1-x)^2 & -x+x^2 \\ x^2 & -x+x^2 & 1-3x+x^2
%  \end{pmatrix}
  \approx \frac{1}{1-4x} 
  \begin{pmatrix}
  1-3x & -x & x^2 \\ -x & 1-2x & -x \\ x^2 & -x & 1-3x
  \end{pmatrix} \; ,
\end{align}
where we neglect the sub-leading $x^2$-terms whenever there is a
linear term as well. The unfolding matrix extends beyond the nearest
neighbor bins, which means that local detector effects lead to a
global unfolding matrix and unfolding only works well if we understand
our entire data set. The reliance on useful kinematic distributions
and the global dependence of the unfolding define the main challenges
once we attempt to unfold the full phase space of an LHC process.

%%%%%%%%%%%%%%%%%%%%%%%%%%%%%%%%%%%%%%%%%%%%%%%%%%%%%%%%%%%%%%%%%%%%%%%%
\subsection{Bayes' theorem and model dependence}
\label{sec:basics_model}

Over the continuous phase space a detector simulation can be written
as
\begin{align}
\frac{\d \sigma}{\d x_d} 
%= \int d x_p \; g(x_d,x_p) \epsilon(x_p) \; \frac{d \sigma}{d x_p} \; ,
= \int \d x_p \; g(x_d,x_p)\; \frac{\d \sigma}{\d x_p} \; ,
\label{eq:toy1}
\end{align}
where $x_d$ is a kinematic variable at detector level, $x_p$ the same
variable at parton level, and $g$ a kernel or transfer function which
links these two arguments. We ignore efficiency factors for now, because
they can be absorbed into the parton-level rate. To invert the
detector simulation we define a second transfer function $\bar{g}$
such that~\cite{Cowan:2002in,Blobel:2011fih,Balasubramanian:2019itp}
\begin{align}
\frac{\d \sigma}{d x_p} 
= \int \d x_d \; \bar{g}(x_p,x_d) \; \frac{\d \sigma}{\d x_d} 
%&= \int d x_d d x_p' \; \bar{g}(x_p,x_d) g(x_p',x_d) \; \frac{d \sigma}{d x_p'} \; ,
= \int \d x_p' \; \frac{\d \sigma}{d x_p'} \int \d x_d \; \bar{g}(x_p,x_d) g(x_d, x_p')  \; .
\label{eq:toy2}
\end{align}
This inversion is fulfilled if we construct the inverse $\bar{g}$ of
$g$ defined by
\begin{align}
\int \d x_d \; \bar{g}(x_p,x_d) g(x_d,x_p') = \delta(x_p - x_p') \; ,
\label{eq:toy3}
\end{align}
all in complete analogy to the binned form above.  The symmetric form
of Eq.\eqref{eq:toy1} and Eq.\eqref{eq:toy2} indicates that $g$ and
$\bar{g}$ are both defined as distributions. In the $g$-direction we
use Monte Carlo simulation and sample in $x_p$, while $\bar{g}$ needs
to be sampled in $g(x_p)$ or $x_d$. In both directions this
statistical nature implies that we should only attempt to unfold
sufficiently large event samples.

The above definitions can be linked to Bayes' theorem if we identify
the kernels with probabilities. We now look at $\bar{g}(x_d|x_p)$
in the slightly modified notation as the probability of observing
$x_d$ given the model prediction $x_p$ and $g(x_p|x_d)$ gives the
probability of the model $x_p$ being true given the observation
$x_d$~\cite{Lucy_1974,Zech_2013}. In this language Eq.\eqref{eq:toy1}
and~\eqref{eq:toy2} describe conditional probabilities, and we can
write something analogous to Bayes' theorem,
\begin{align}
  \bar{g}(x_p|x_d) \; \frac{\d \sigma}{\d x_d}
  \sim g(x_d|x_p) \; \frac{\d \sigma}{\d x_p} \; .
\end{align}
In this form $\bar{g}(x_p|x_d)$ is the posterior, $g(x_d|x_p)$ as a
function of $x_p$ is the likelihood, $\d \sigma/\d x_p$ is the prior,
and the model evidence $\d \sigma/\d x_d$ fixes the normalization of the
posterior.  From standard Bayesian analyses we know two things: (i)
the posterior will in general depend on the prior, in our case the
kinematics of the underlying particle physics process or model; (ii)
when analyzing high-dimensional spaces the prior dependence will
vanish when the likelihood develops a narrow global maximum.

If the posterior $\bar{g}(x_p|x_d)$ in general depends on the model $\d
\sigma/\d x_p$, then Eq.\eqref{eq:toy2} does not look useful. On the
other hand, Bayesian statistics is based on the assumption that the
prior dependence of the posterior defines an iterative process where
we start from a very general prior and enter likelihood information
step by step to finally converge on the posterior. The same approach
can define a kinematic unfolding
algorithm~\cite{DAgostini:1994fjx}. We will not discuss these
methods further, but come back to this model dependence throughout our
paper.

%%%%%%%%%%%%%%%%%%%%%%%%%%%%%%%%%%%%%%%%%%%%%%%%%%%%%%%%%%%%%%%%%%%%%%%%
\subsection{Reference process $pp \to ZW$}
\label{sec:basics_proc}
%-------------------------------------------------
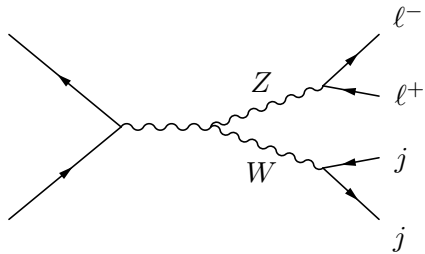
\begin{figure}[b!]
\begin{center}
\input{incl_feynman}
\end{center}
\caption{Sample Feynman diagram contributing to $ZW$ production, with
  intermediate on-shell particles labeled.}
\label{fig:feyn_intro}
\end{figure}
%------------------------------------------------------------

To provide a quantitative estimate of unfolding with an invertible
neural networks we use the same example process as in
Ref.~\cite{fcgan},
\begin{align}
pp
\to ZW^\pm
\to (\ell^- \ell^+) \; (j j ) \; ,
\end{align}
One of the contributing Feynman diagrams is shown in
Fig.~\ref{fig:feyn_intro}. With jets and leptons in the final state we
can test the stability of the unfolding network for small and for
large detector effects.
%In Sec.~\ref{sec:jets_model} we will also
%introduce a $W'$ resonance in the $s$-channel.
We generate the $ZW$ events using \madgraph~\cite{madgraph} without
any generation cuts and then simulate parton showering with
\pythia~\cite{pythia8} and the detector effects with
\delphes~\cite{delphes} using the standard ATLAS card.  For jet
clustering we use the anti-$k_T$ algorithm~\cite{anti-kt} with
$R=0.6$ implemented in \fastjet~\cite{fastjet}. All jets are required
to have
\begin{align}
p_{T, j} > 25~\gev 
\qquad \text{and} \qquad 
|\eta_j| < 2.5 \; .
\label{eq:jetcond}
\end{align}
For the hadronically decaying $W$-boson the limited calorimeter
resolution will completely dominate over the parton-level Breit-Wigner
distribution.  After applying the cuts we have 320k events which we
split into 90\% training and 10\% test data.

In a first step, corresponding to Ref.~\cite{fcgan} we are only
interested in inverting these detector effects. These results are
shown in Sec.~\ref{sec:inn}. For the simulation this implies that we
switch off initial state radiation as well as underlying event and
pile-up effects and require exactly two jets and a pair of same-flavor
opposite-sign leptons. The jets and corresponding partons are
separately ordered by $p_T$. The detector and parton level leptons are
assigned by charge. This gives us two samples matched event by event,
one at the parton level ($x_p$) and one including detector effects
($x_d$).  Each of them is given as an unweighted set of four
4-vectors. These 4-vectors can be simplified if we assume all external
particles at the parton level to be on-shell. Obviously, this
method can be easily adapted to weighted events.

In a second step we include initial state radiation and allow for
additional jets in Sec.~\ref{sec:jets}.  We still require a pair of
same-flavor opposite-sign leptons and at least two jets in agreement
with the condition in Eq.\eqref{eq:jetcond}. The four jets with
highest $p_T$ are then used as input to the network, ordered by
$p_T$. Events with less than 4 jets are zero-padded. This second
data set is only used for the conditional INN.

%%%%%%%%%%%%%%%%%%%%%%%%%%%%%%%%%%%%%%%%%%%%%%%%%%%%%%%%%
\section{Unfolding detector effects}
\label{sec:inn}

We introduce the conditional INN in two steps, starting with the
non-conditional, standard setup. The construction of the INN we use in
our analysis combines two goals~\cite{inn1}:
\begin{enumerate}
\item the mapping from input to output is invertible and the Jacobians
  for both directions are tractable;
\item both directions can be evaluated efficiently. This second
  property goes beyond some other implementations of normalizing
  flow~\cite{nflow1,nflow_review}.
\end{enumerate}
While the final aim is not actually to evaluate our INN in both directions, we will
see that these networks can be extremely useful to invert a Markov
process like detector smearing. Their bi-directional training makes
them especially stable.

In Sec.~\ref{sec:inn_cond} we will show how the conditional INN
retains a proper statistical notion of the inversion to parton level
phase space.  This avoids a major weakness of standard unfolding
methods, namely that they only work on large enough event samples
condensed to one-dimensional or two-dimensional kinematic
distributions. This could be a missing transverse energy distribution
in mono-jet searches or the rapidities and transverse momenta in top
pair production. To avoid systematics or biases in the full phase
space coverage required by the matrix element method, the unfolding
needs to construct probability distributions in parton-level phase
space, including small numbers of events in tails of kinematic
distributions.

%%%%%%%%%%%%%%%%%%%%%%%%%%%%%%%%%%%%%%%%%%%%%%%%%%%%%%%%%
\subsection{Naive INN}
\label{sec:inn_base}

%------------------------------------------------------------
\begin{figure}[t]
\centering
\input{incl_network_inn}
\caption{Structure of INN. The $\{ x_{d,p} \}$ denote detector-level
  and parton-level events, $\{ r_{d,p} \}$ are random numbers to match
  the phase space dimensionality. A tilde indicates the INN
  generation.}
\label{fig:inn}
\end{figure}
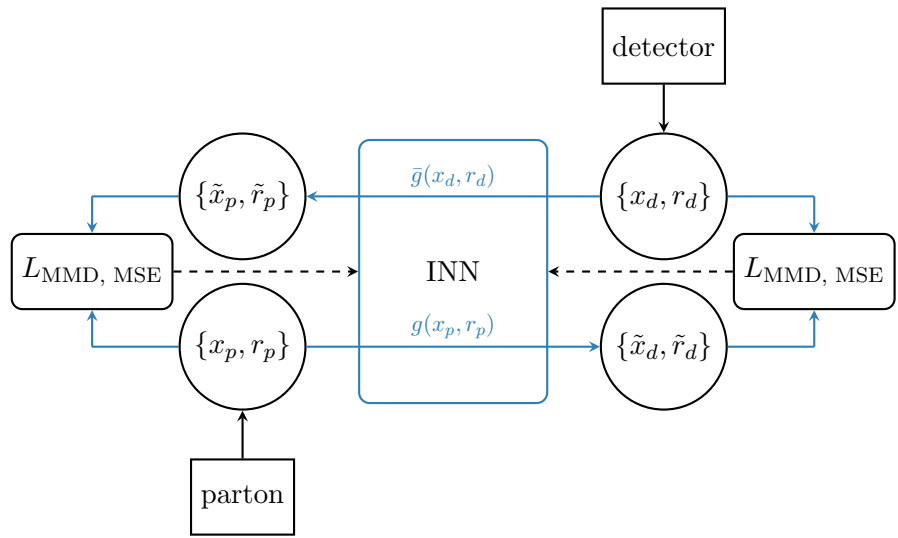
%------------------------------------------------------------

While it is clear from our discussion in Ref.~\cite{fcgan} that a
standard INN will not serve our purpose, we still describe it in some
detail before we extend it to a conditional network.  Following the
conventions of our GAN analysis and in analogy to Eqs.\eqref{eq:toy1}
to \eqref{eq:toy3} we define the network input as a vector of hard
process information $x_p \in R^{D_p}$ and the output at detector level
via the vector $x_d \in \mathbb{R}^{D_d}$. If the dimensionalities of the
spaces are such that $D_p < D_d$ we add a noise vector $r$ with dimension
$D_d-D_p$ to define the bijective, invertible transformation,
\begin{align}
\begin{pmatrix} x_p \\ r \end{pmatrix}
\stackrel[\leftarrow \; \text{unfolding}: \bar{g}]{\textsc{Pythia,Delphes}: g \rightarrow}{\xleftrightarrow{\hspace*{3.5cm}}}
 x_d  \; .
\label{eq:mapping}
\end{align}
A correctly trained network $g$ with the parameters $\theta$
then reproduces $x_d$ from the combination $x_p$ and $r$. Its inverse
$\bar{g}$ instead reproduces the combination of $x_p$ and $r$ from $x_d$.

The defining feature of the INN illustrated in Fig.~\ref{fig:inn} is
that it learns both directions of the bijective mapping in parallel
and encodes them into one network. Such a simultaneous training of
both directions is guaranteed by the building blocks of the network,
the invertible coupling layers~\cite{coupling1,coupling2}.  For
notational purposes we ignore the random numbers in
Eq.\eqref{eq:mapping} and assume that this layer links an input vector
$x_p$ to an output vector $x_d$ after splitting both of them in
halves, $x_{p,i}$ and $x_{d,i}$ for $i=1,2$. The relation between
input and output is given by a sub-network, which encodes arbitrary
functions $s_{1,2}$ and $t_{1,2}$.  Using an element-wise
multiplication~$\odot$ and sum one could for instance define an output
$x_{d,1}(x_p) = x_{p,1} \odot s_2(x_{p,2}) + t_2(x_{p,2})$. In order to avoid numerical instabilities caused by the division with $s(x)$ in the inverse direction, we include an exponential to obtain
\begin{align}
\begin{pmatrix} x_{d,1} \\ x_{d,2} \end{pmatrix} = 
\begin{pmatrix}
x_{p,1} \odot e^{s_2(x_{p,2})} + t_2(x_{p,2}) \\
x_{p,2} \odot e^{s_1(x_{d,1})} + t_1(x_{d,1}) 
\end{pmatrix}
\hspace{0.5em} \Leftrightarrow \hspace{0.5em}
\begin{pmatrix} x_{p,1} \\ x_{p,2} \end{pmatrix} = 
\begin{pmatrix}
(x_{d,1} - t_2(x_{p,2})) \odot e^{-s_2(x_{p,2})} \\
(x_{d,2} - t_1(x_{d,1})) \odot e^{-s_1(x_{d,1})} 
\end{pmatrix} \; .
\label{eq:layers}
\end{align}
By construction, this inversion works independent of the form of $s$
and $t$. If we write the coupling block function as $g(x_p) \sim x_d$,
again omitting the random numbers $r$, the Jacobian of the network
function has a triangular form
\begin{align}
\frac{\partial g(x_p)}{\partial x_p} = 
\begin{pmatrix}
\text{diag } e^{s_2(x_{p,2})} & \text{finite} \\
0 & \text{diag } e^{s_1(x_{d,1})}
\end{pmatrix} \; ,
\label{eq:jacob}
\end{align}
so its determinant is easy to compute. Such coupling layer
transformations define a so-called normalizing flow, when we view it
as transforming an initial probability density into a very general
form of probability density through a series of invertible steps. We
can relate the two probability densities as long as the Jacobians of
the individual layers can be efficiently calculated.

Since the first use of the invertible coupling layer, much effort has
gone into improving its efficiency. The All-in-One (AIO) coupling
layer includes two features, introduced by Ref.~\cite{coupling2} and
Ref.~\cite{glow}. The first modification replaces the transformation
of $x_{p,2}$ by a permutation of the output of each layer. Due to the
permutation each component still gets modified after passing through
several layers. The second modification includes a global affine transformation
to include a global bias and linear scaling that maps $x \rightarrow s
x + b$. Finally, we apply a bijective soft clamping after the
exponential function in Eq.\eqref{eq:layers} to prevent instabilities
from diverging outputs.

%In a simplified approach we will first use pairs of events with
%matching numbers of objects for input and output, represented by
%four-vectors. This approach neglects the difference in degrees of
%freedom and the statistical nature of the process, but it will allow
%us to demonstrate the efficiency of the the INNs.

The INN in our simplified example combines three contributions to the
loss function. First, it tests if in the \textsc{Delphes} direction of
Eq.\eqref{eq:mapping} we indeed find $g(x_p) = x_d$ via the mean
squared error (MSE) function. While this is theoretically sufficient
to obtain the inverse function, also testing the inverse direction
$\bar{g}(x_d) = x_p$ greatly improves the efficiency and stability of
the training. Third, to resolve special sharp features like the
invariant mass of intermediate particles we use the maximum mean
discrepancy (MMD) as a distance measure between the generated and real
distribution of these features.

Because we will also use the MMD in another function
function~\cite{gan_phasespace} we review it briefly. An MMD loss
allows us to compare any pre-defined distribution. For a relativistic
phase space a critical narrow phase space feature is the invariant
mass of intermediate particles. We can force the network to consider
this one-dimensional distribution of the 4-vectors $x_p$ for batches
of parton-level and detector-level events,
\begin{align}
\text{MMD} =
\left[ \langle k\left(x,x'\right)\rangle_{x,x' \sim P_p} 
     + \langle k\left(y,y'\right)\rangle_{y,y' \sim P_d} 
     - 2 \langle k\left(x,y \right)\rangle_{x\sim P_p,y \sim P_d} \right]^{1/2} \; .
\label{eq:MMD}
\end{align}
In Refs.~\cite{gan_phasespace} and~\cite{fcgan} we compare common
choices, like Gaussian or Breit-Wigner kernels
\begin{align}
k_\text{Gauss} \left(x,y\right) = \exp \frac{- \left(x - y\right)^2}{2 \sigma^2}
\qquad \text{or} \qquad 
k_\text{BW}\left(x,y\right) = \frac{\sigma^2}{\left(x - y\right)^2 + \sigma^2} 
\label{eq:kernels}
 \end{align}
with a fixed or variable width $\sigma$~\cite{fcgan}.  Inside the INN
architecture the Breit-Wigner kernel is the best choice to analyze the
distribution of the random numbers as part of the loss
function~\cite{inn1}.

%------------------------------------------------------------
\begin{figure}[t]
%\centering
\includegraphics[page=25, width=0.5\textwidth]{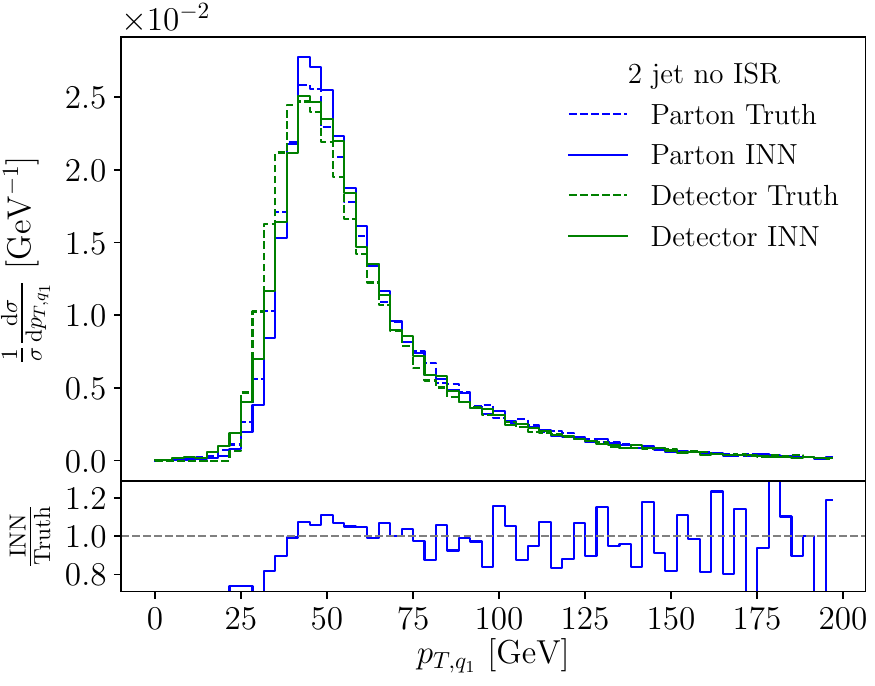}
\includegraphics[page=25, width=0.5\textwidth]{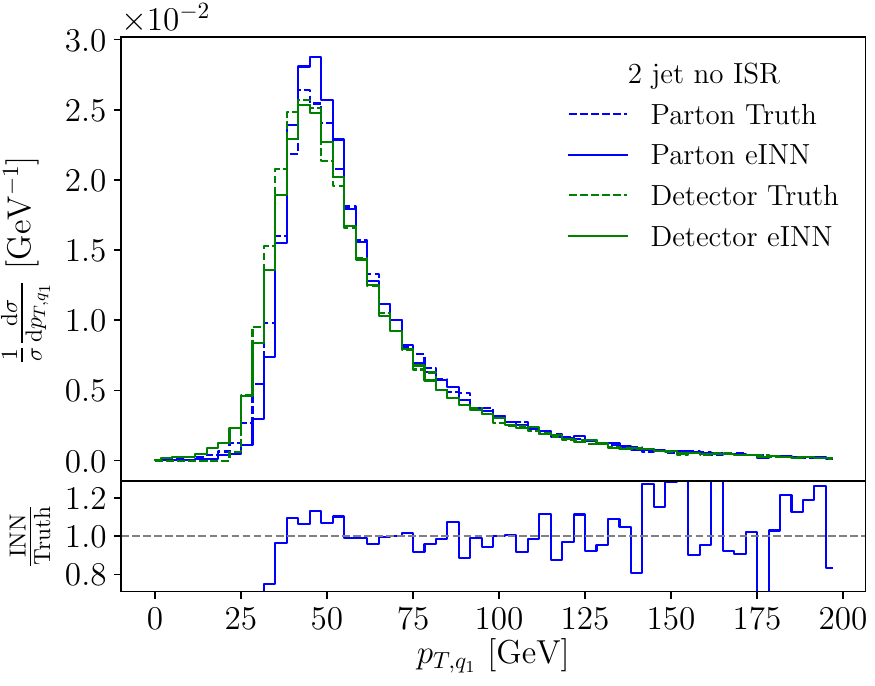} \\
\includegraphics[page=26, width=0.5\textwidth]{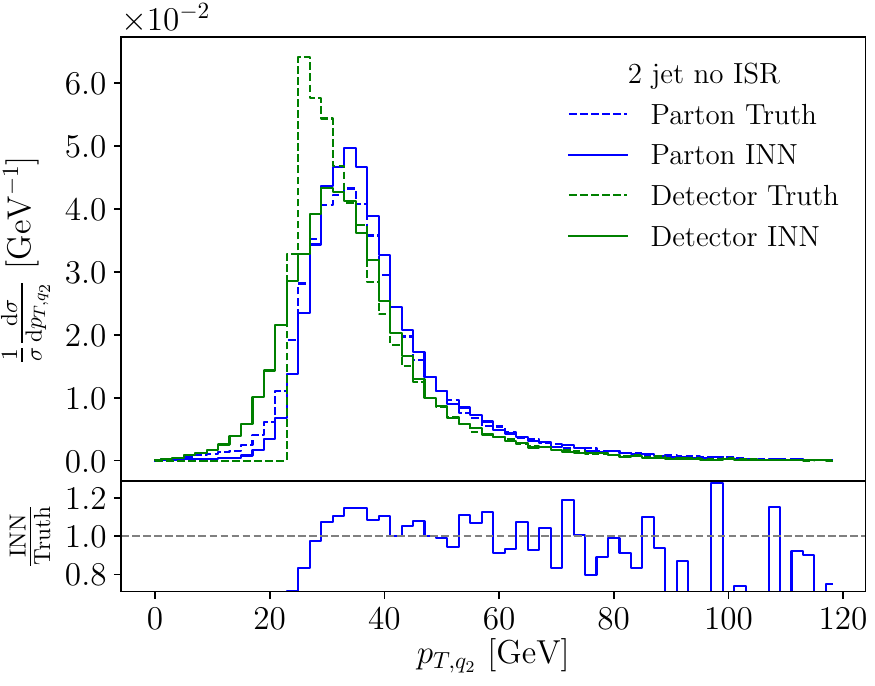}
\includegraphics[page=26, width=0.5\textwidth]{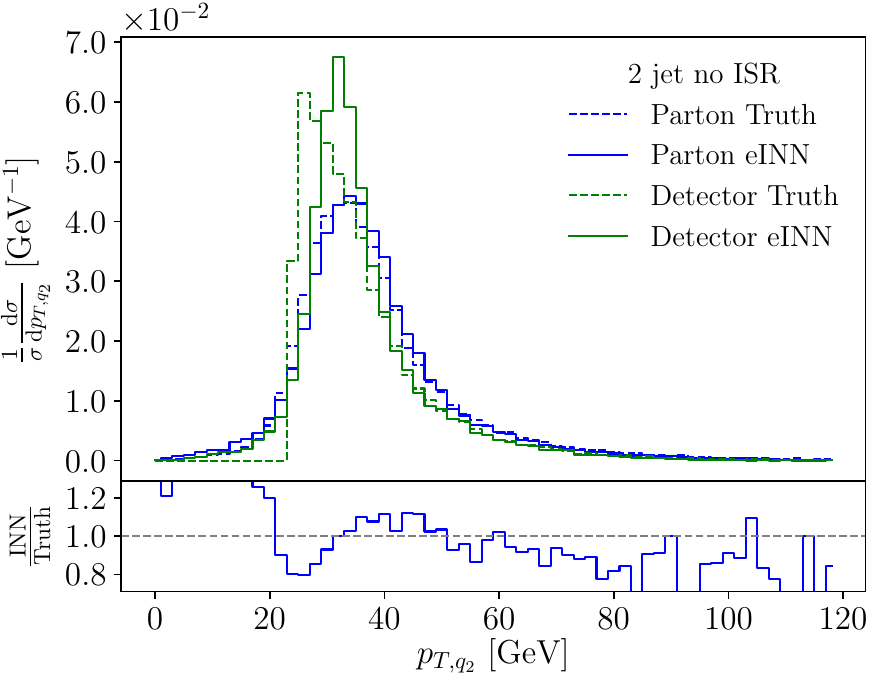} \\
\includegraphics[page=32, width=0.5\textwidth]{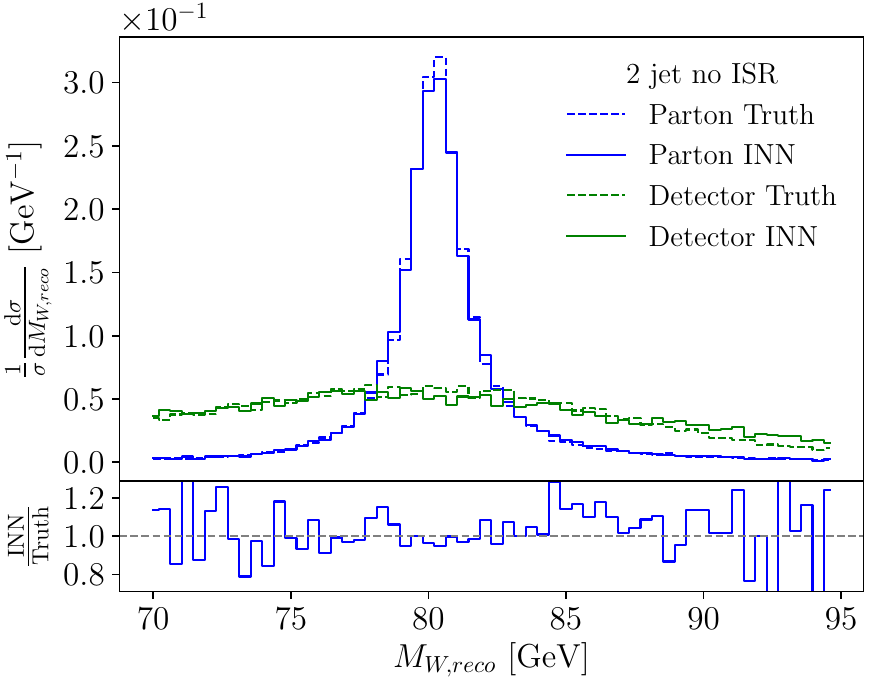}
\includegraphics[page=32, width=0.5\textwidth]{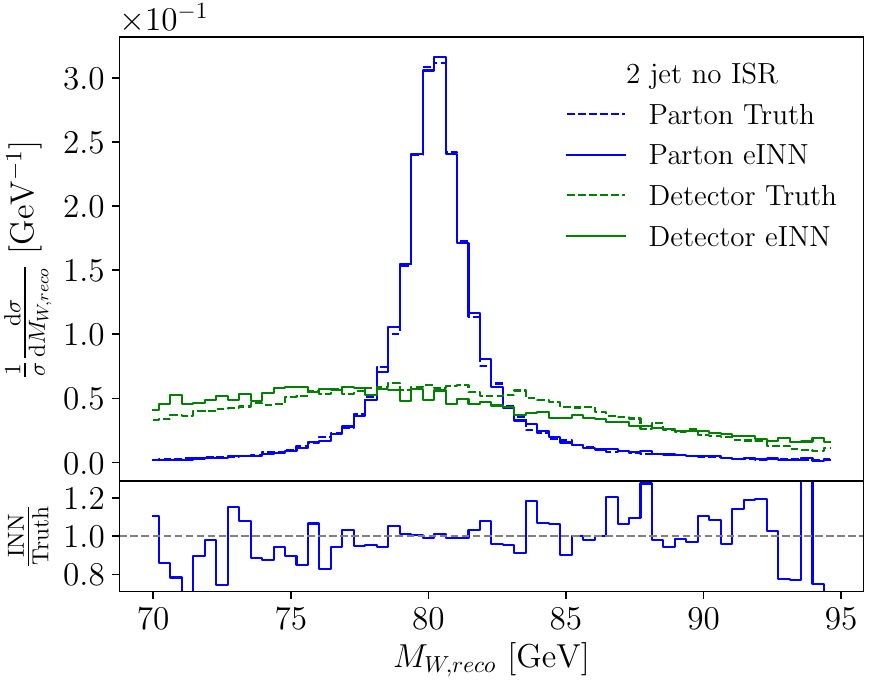}
\caption{INNed $p_{T,q}$ and $M_{W,\text{reco}}$ distributions from a
  naive INN (left) and the noise-extended eINN (right). In green we
  compare the detector-level truth to INNed events transformed from
  parton level. In blue we compare the parton-level truth to INNed
  events transformed from detector level. The secondary panels show
  the ratio of INNed events over parton-level truth. More
  distributions can be found in the pdf files submitted to the arXiv.}
\label{fig:UnfoldCurve}
\end{figure}
%------------------------------------------------------------

We now use the INN network to map parton-level events to
detector-level events or vice-versa. In a statistical analysis we then
use standard kinematic distributions and compare the respective truth
and INN-inverted shapes for both directions. The left panels of
Fig.~\ref{fig:UnfoldCurve} shows the transverse momentum distributions
of the two jets and their invariant mass for both directions of the
INN. The truth events at parton level and at detector level are marked
as dashed lines. Starting from each of the truth events we can apply
the INN describing the detector effects as $x_d = g(x_p)$ or unfolding
the detector effects as $x_p = \bar{g}(x_d)$ in
Eq.\eqref{eq:mapping}. The corresponding solid lines have to be
compared to the dotted truth lines, where we need to keep in mind that
at the parton level the relevant objects are quarks while at the
detector level they are jets.

For the leading jet the truth and INNed detector-level agree very
well, while for the second jet the naive INN fails to capture the hard
cut imposed by the jet definition. For the invariant mass we find that
the smearing due to the detector effects is reproduced well with some
small deviations in the tails. In the unfolding direction both $p_T$
distributions follow the parton level truth. The only difference is a
systematic lack of events in the tail for the second quark. This is
especially visible in the ratio of the INN-unfolded events and the
parton-level truth, indicating that also at small $p_T$ the network
does not fill the phase space sufficiently. Combining both directions
we see that in forward direction the INN produces a too broad
$p_T$-distribution, the unfolding direction of the INN produces a too
narrow distribution.  The conceptual advantage of the INN actually
implies a disadvantage for the inversion of particular difficult
features.  Finally, the invariant mass of the $W$ is reproduced
perfectly without any systematic deviation.

%%%%%%%%%%%%%%%%%%%%%%%%%%%%%%%%%%%%%%%%%%%%%%%%%%%%%%%%%
\subsection{Noise-extended INN}
\label{sec:inn_noise}

While our simplified example in the previous section shows some serious
promise of INNs, it fails to incorporate key aspects of the physical
process.  First of all, the number of degrees of freedom is not
actually the same at parton level and at detector level. External
partons are on their mass shell, while jets come with a range of jet
masses. This mismatch becomes crucial when we include missing
transverse momentum in the signature.  We generally need fewer
parameters to describe the partonic scattering than the detector-level
process.  For a fixed set of parton-level momenta we usually smear
each momentum component to simulate the detector measurement. These
additional degrees of freedom are of stochastic nature, so adding
Gaussian random variable on the parton side of the INN could be a
first step to address this problem.

To also account for potentially unobservable degrees of freedom at the
parton level we extend each side of the INN by a random number vector.
The mapping in Eq.\eqref{eq:mapping} now includes two random number
vectors with dimensions $D_{r_d} = D_p$ and $D_{r_p} = D_d$,
\begin{align}
\begin{pmatrix} x_p \\ r_p \end{pmatrix}
\stackrel[\leftarrow \; \text{unfolding}: \bar{g}]{\textsc{Pythia,Delphes}: g \rightarrow}{\xleftrightarrow{\hspace*{3.5cm}}}
\begin{pmatrix} x_d \\ r_d \end{pmatrix} \; .
\label{eq:mappingnoise}
\end{align}
In addition, a pure MSE loss can not capture the fact that the
additional noise generates a distribution of detector-level events
given fixed parton momenta. It would just predict of a mean value of
this distribution and minimize the effect of the noise. A better
solution is an MMD loss for each degree of freedom in the event and
the masses of intermediate particles, as well as the Gaussian random
variables. On the side of the random numbers this MMD loss ensures
that they really only encode noise. Again it is beneficial for the
training to use the inverse direction and apply additional MMD
losses to the parton level events as well as the corresponding
Gaussian inputs.  Finally we add a weak MSE loss on the four vectors
of each side to stabilize the training.

In the right panels of Fig.~\ref{fig:UnfoldCurve} we show results for
this noise-extended INN (eINN). The generated distributions are similar to
the naive INN case and match the truth at the parton level. A notable
difference appears in the second jet, the weak spot of the naive
INN. The additional random numbers and MMDs provide more freedom to
generate the peak in the forward direction and also improve the
unfolding in the low-$p_T$ and high-$p_T$ regimes.\bigskip

%------------------------------------------------------------
\begin{figure}[t]
\includegraphics[page=9, width=0.505\textwidth]{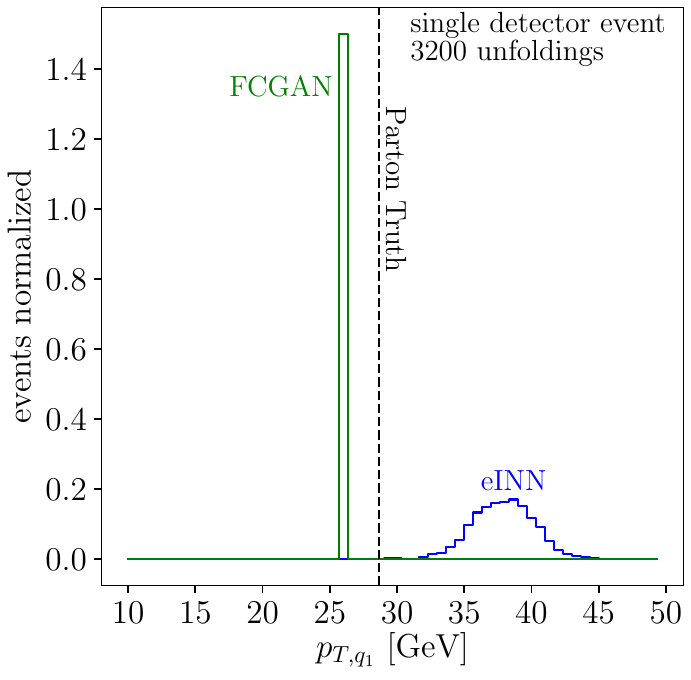}
\includegraphics[page=20, width=0.495\textwidth]{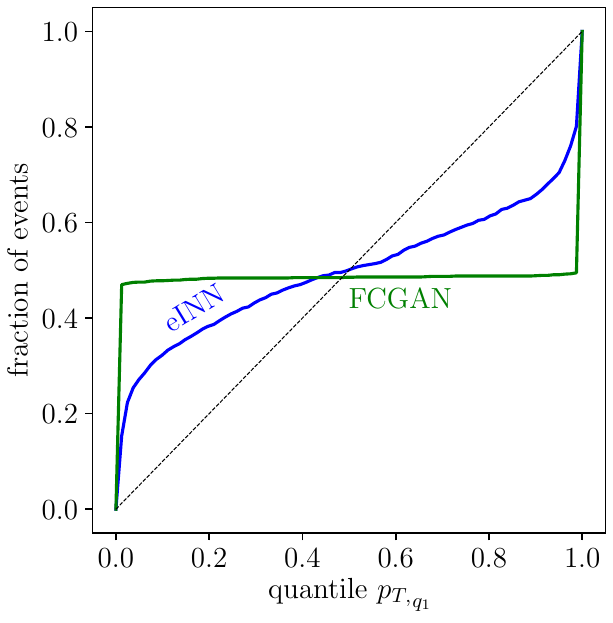}
\caption{Left: illustration of the statistical interpretation of
  unfolded events for one event. Right: calibration curves for
  $p_{T,q_1}$ extracted from the FCGAN and the noise-extended eINN.}
\label{fig:quantile1}
\end{figure}
%------------------------------------------------------------

Aside from the better modeling, the noise extension allows for a
statistic interpretation of the generated distributions and a test of
the integrity of the INN-inverted distributions. In the left panel of
Fig.~\ref{fig:quantile1} we illustrate the goal of the statistical
treatment: we start from a single event at the detector level and
generate a set of unfolded events. For each of them we evaluate for
instance $p_{T,q_1}$. Already in this illustration we see that the GAN
output is lacking a statistical behavior at the level of individual
events, while the noise-extended eINN returns a reasonable
distribution of unfolded events.

To see if the width of this INN output is correct we take 1500
parton-level and detector-level event pairs and unfold each event 60
times, sampling over the random variables. This gives us 1500
combinations like the one shown in the left panel of
Fig.~\ref{fig:quantile1}: a single parton-level truth configuration
and a distribution of the INNed configuration. To see if the central
value and the width of the INNed distribution can be interpreted
statistically as a posterior probability distribution in parton phase
space we analyse where the truth lies within the INN distribution for
each of the 1500 events.  For a correctly calibrated curve we start
for instance from the left of the kinematic distribution and expect
10\% of the 1500 events in the 10\% quantile of the respective
probability distribution, 20\% of events in the 20\% quantile, etc.
The corresponding calibration curves for the noise-extended eINN are
shown in the right panel of Fig.~\ref{fig:quantile1}. While they
indicate that we can attempt a statistical interpretation of the INN
unfolding, the calibration is not (yet) perfect.  A steep rise for the
lower quantile indicates that too many events end up in the first 10\%
quantile. In other words, the distributions we obtain by sampling over
the Gaussian noise for each event are too narrow.

While our noise-extended eINN takes several steps in the right
direction, it still faces major challenges: the combination of many
different loss functions is sensitive to their relative weights; the
balance between MSE and MMD on event constituents has to be calibrated
carefully to generate reasonable quantile distributions; when we want
to extend the INN to include more detector-level information we have
to include an equally large number of random variable on the parton
level which makes the training very inefficient. This leads us
again~\cite{fcgan} to adopt a conditional setup.

%%%%%%%%%%%%%%%%%%%%%%%%%%%%%%%%%%%%%%%%%%%%%%%%%%%%%%%%%%%%%%%%%%%%%%%%
\subsection{Conditional INN}
\label{sec:inn_cond}

If a distribution of parton-level events can be described by $n$
degrees of freedom, we should be able to use normalizing flows or an
INN to map a $n$-dimensional random number vector onto parton-level
4-momenta.  To capture the information from the detector-level events
we need to condition the INN on these
events~\cite{goodfellow,cond_gan,fcgan}, so we link the parton-level
data $x_p$ to random noise $r$ under the condition of $x_d$. Trained
on a given process the network should now be able to generate
probability distributions for parton-level configurations given a
detector-level event and an unfolding model. We note that the cINN
is still invertible in the sense that it includes a bi-directional
training from Gaussian random numbers to parton-level events and
back. While this bi-directional training does not represent the
inversion of a detector simulation anymore, it does stabilize the
training by requiring the noise to be Gaussian.

%------------------------------------------------------------
\begin{figure}[t]
\centering
\input{incl_network_cinn}
\caption{Structure of the conditional INN. The input are random
  numbers $\{ r\}$ while $\{ x_{d,p} \}$ denote detector-level and
  parton-level data. The latent dimension loss $L$ follows
  Eq.\eqref{eq:loss}, a tilde indicates the INN generation.}
\label{fig:cinn}
\end{figure}
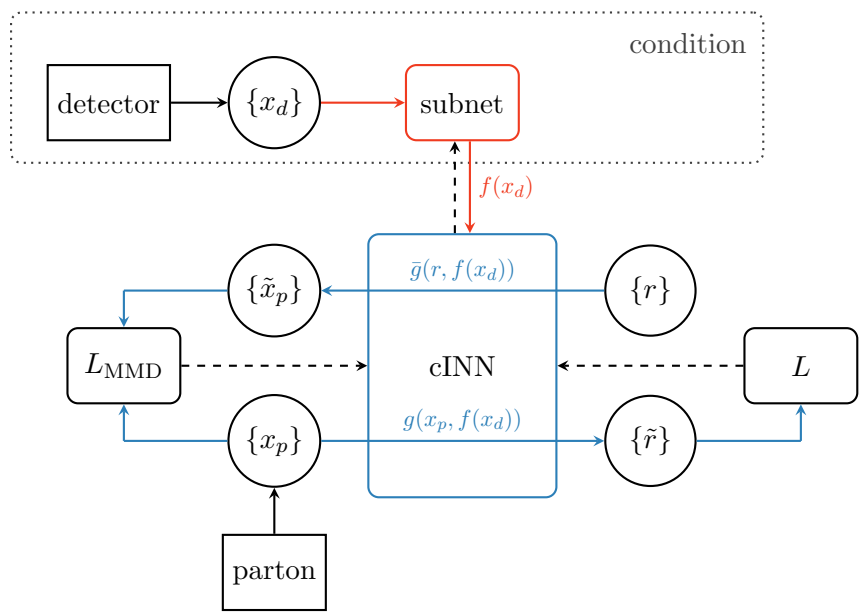
%------------------------------------------------------------

%------------------------------------------------------------
\begin{table}[b!]
\centering
\begin{small} \begin{tabular}{l|c c}
\toprule
Parameter & INN & eINN  \\
\midrule
Blocks & 24 & 24\\
Layers per block & 2 & 2\\
Units per layer & 256 & 256\\
Trainable weights & $\sim$ 150k & $\sim$ 270k \\
Epochs & 1000 & 1000 \\
Learning rate & $8 \cdot 10 ^{-4}$ & $8 \cdot 10 ^{-4}$\\
Batch size & 512 & 512 \\
Training/testing events & 290k / 30k & 290k / 30k \\
Kernel widths & $\sim 2, 8, 25, 67$ & $\sim 2, 8, 25, 67$\\
$D_p+D_{r_p}$ & $12+4$ & $12+16$ \\
$D_d+D_{r_d}$ & $16+0$ & $16+12$ \\
$\lambda_\text{MMD}$ & 0.1 (masses only) & 0.2 \\
$\lambda_\text{MMD}$ increase & - & - \\
\bottomrule
\end{tabular} \end{small}
\caption{INN and noise-extended eINN setup and hyper-parameters, as
  implemented in \pytorch(v1.2.0)~\cite{pytorch}.}
\label{tab:inn}
\end{table}
%------------------------------------------------------------

%------------------------------------------------------------
\begin{figure}[t]
\includegraphics[page = 2, width=0.5\textwidth]{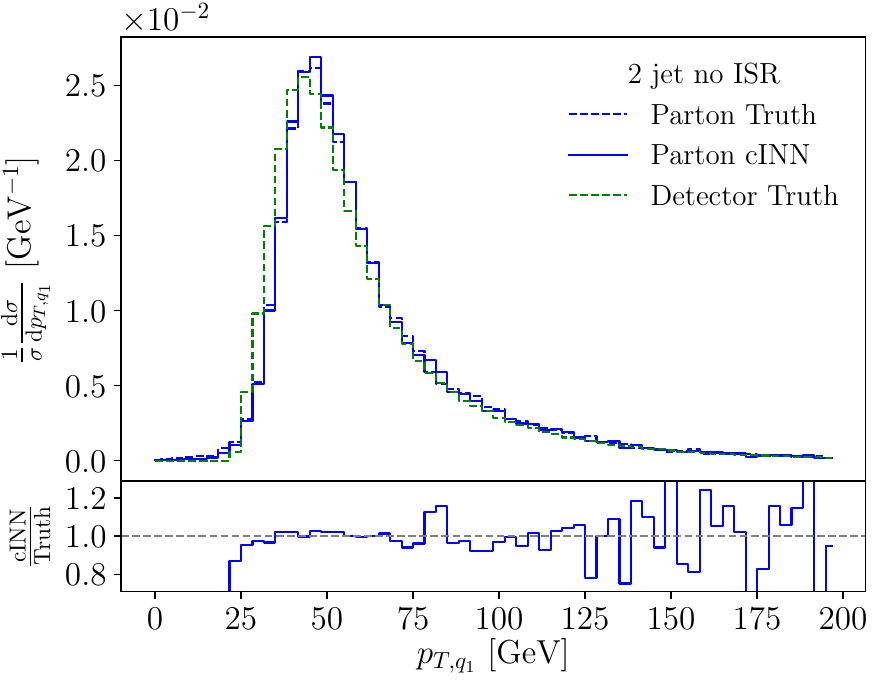}
\includegraphics[page = 9, width=0.5\textwidth]{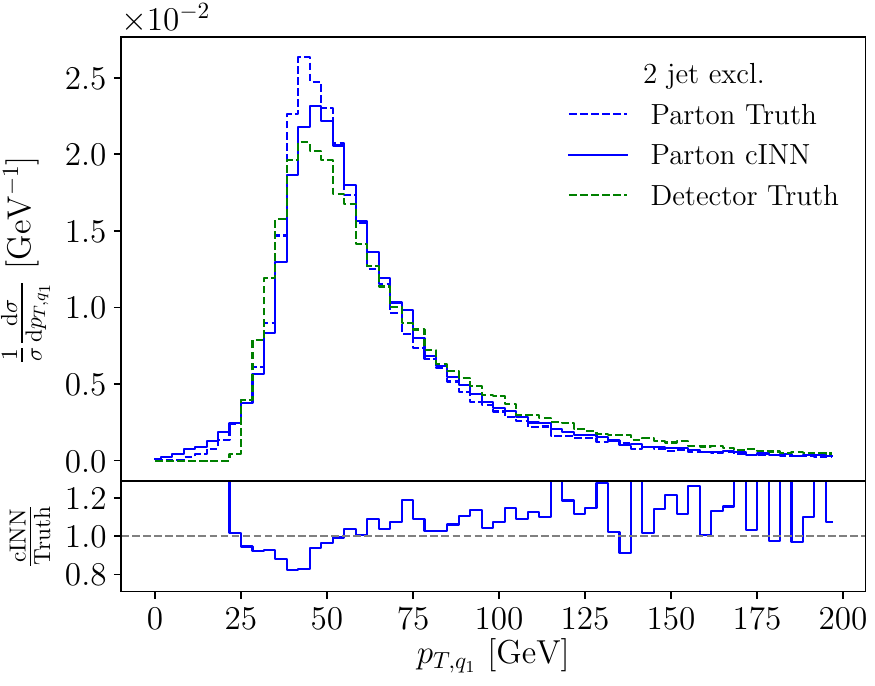} \\
\includegraphics[page =3, width=0.5\textwidth]{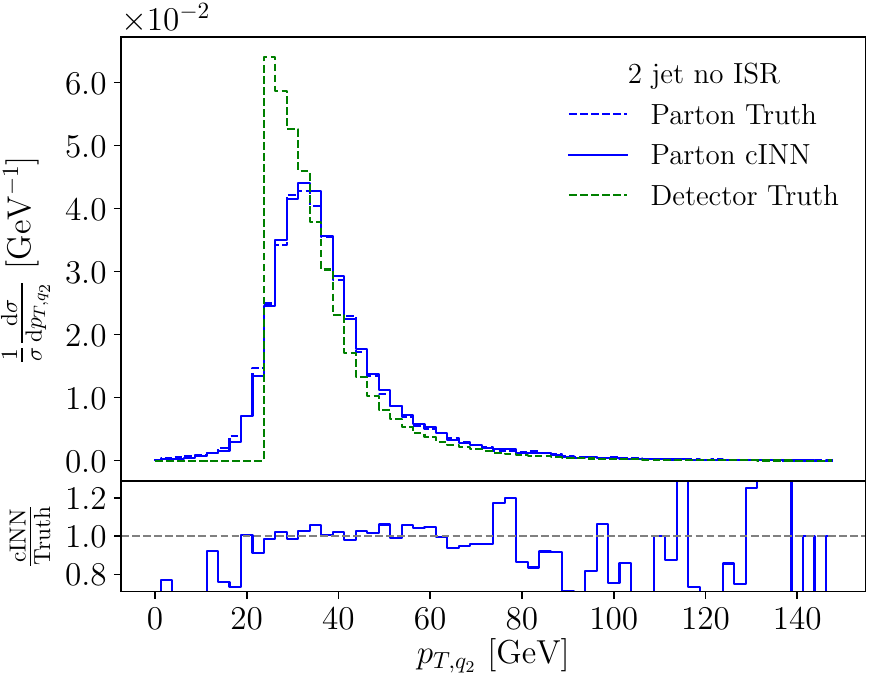} 
\includegraphics[page =10, width=0.5\textwidth]{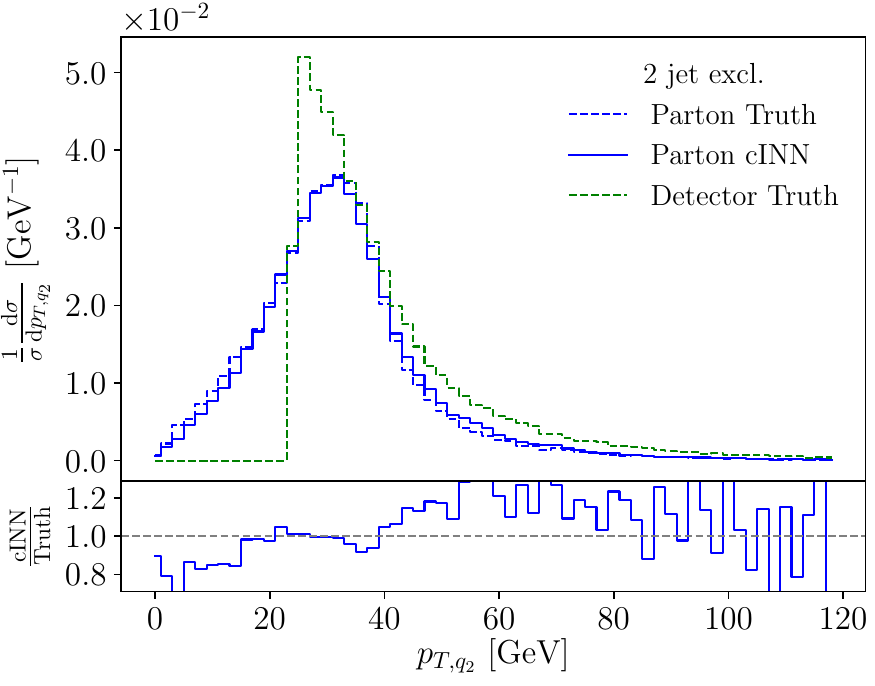} \\
\includegraphics[page =4, width=0.5\textwidth]{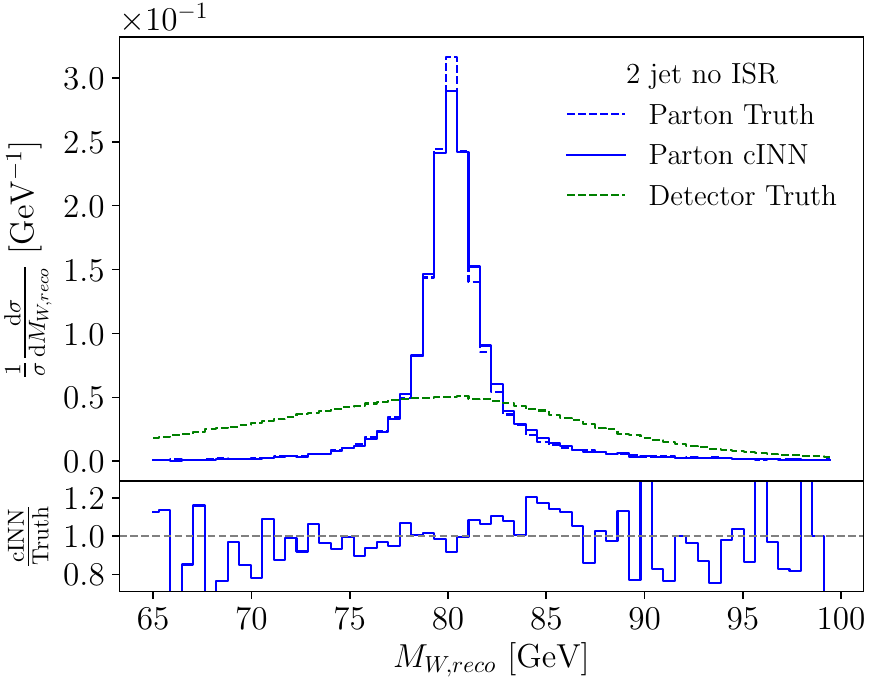}
\includegraphics[page =19, width=0.5\textwidth]{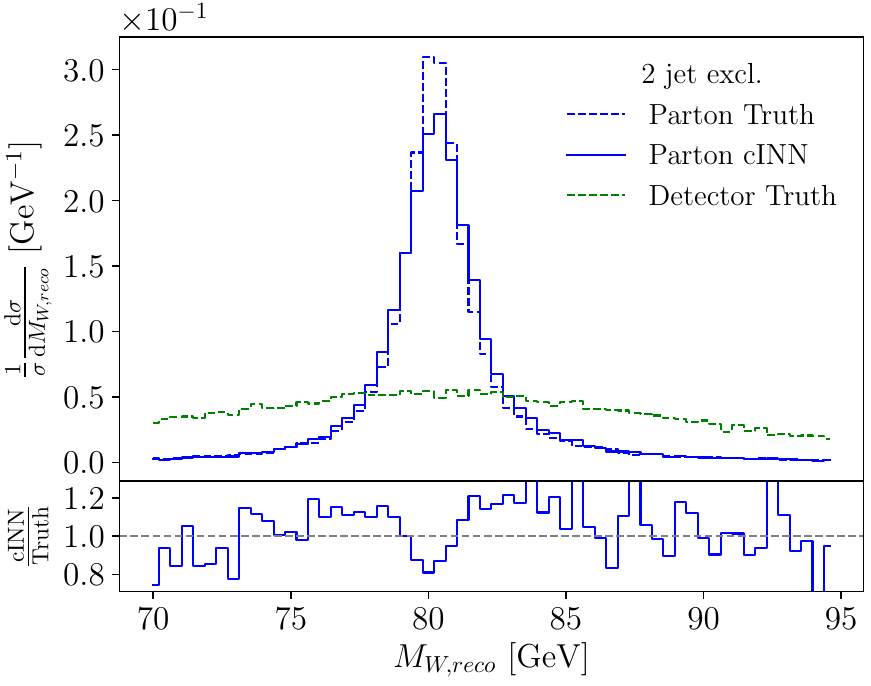}
\caption{cINNed $p_{T,q}$ and $m_{W,\text{reco}}$ distributions.  Training and
  testing events include exactly two jets. In the left panels we use a
  data set without ISR, while in the right panels we use the two-jet
  events in the full data set with ISR. The lower panels give the
  ratio of cINNed to parton-level truth.}
\label{fig:2j}
\end{figure}
%------------------------------------------------------------

A graphic representation of this conditional INN or cINN is given in
Fig.~\ref{fig:cinn}.  We first process the detector-level data by a
small subnet, \ie $x_d\to f(x_d)$, to optimize its usability for the cINN~\cite{cinn}. The
subnet is trained alongside the cINN and does not need to be reversed
or adapted.  We choose a shallow and wide architecture of two layers
with a width of 1024 internally, because four layers
degrade already the conditional information and allow the cINN to ignore it.
When a deeper subnet is required we advertize to use an
encoder, which is initialized by pre-training it as part of an autoencoder. We apply this technique when using the larger ISR input, where it leads to a more efficient training.  After this preprocessing, the detector information is passed
to the functions $s_i$ and $t_i$ in Eq.\eqref{eq:layers}, which now
depend on the input, the output, and on the fixed condition. Since the
invertibility of the network is independent of the values of $s_i$ and
$t_i$, the network remains invertible between the parton-level events
$\{ x_p \}$ and the random variables $\{ r \}$.  This feature
stabilizes the training. The cINN loss function is motivated by the
simple argument that for the correct set of network parameters
$\theta$ describing $s_i$ and $t_i$ we maximize the (posterior)
probability $p(\theta |x_p,x_d)$ or minimize
%
%\begin{align}
%L &= - \mathbb{E} \left[ \log p(\theta |x_p,x_d) \right] \notag \\
%  &\approx - \mathbb{E} \left[ \log p(x_d |x_p,\theta) \right] - \log p(\theta) \notag \\
%% &= - \mathbb{E} \left[ \log p(g(x_p) |x_p,\theta) + \log \left| \frac{\partial g}{\partial x_p} \right| \right] - \log p(\theta) \notag \\
%  &= - \mathbb{E} \left[ \log p(g(x_p)) + \log \left| \frac{\partial g}{\partial x_p} \right| \right] - \log p(\theta) \; ,
%\label{eq:loss}
%\end{align}
\begin{align}
%\begin{split}
%L &= -  \left\langle \log p(\theta |x_p,x_d) \right\rangle_{x_p\sim P_p,x_d \sim P_d} \\
%&= - \left\langle  \log p(x_d |x_p,\theta) \right\rangle_{x_p\sim P_p,x_d \sim P_d}  - \log p(\theta) + \text{const.} \\
%&= - \left\langle \log p(g(x_p,f(x_d))) + \log \left| \frac{\partial g(x_p,f(x_d))}{\partial x_p} \right| \right\rangle_{x_p\sim P_p,x_d \sim P_d}  - \log p(\theta) + \text{const.} \; ,
%\end{split}
\begin{split}
L &= -  \left\langle \log p(\theta |x_p,x_d) \right\rangle_{x_p\sim P_p,x_d \sim P_d} \\
&= -  \left\langle \log p(x_d |x_p, \theta) + \log p(\theta|x_p) - \log p(x_d|x_p)\right\rangle_{x_p\sim P_p,x_d \sim P_d} \\
&= - \left\langle  \log p(x_d |x_p,\theta) \right\rangle_{x_p\sim P_p,x_d \sim P_d}  - \log p(\theta) + \text{const.} \\
&= - \left\langle \log p(g(x_p,x_d)) + \log \left| \frac{\partial g(x_p,x_d)}{\partial x_p} \right| \right\rangle_{x_p\sim P_p,x_d \sim P_d}  - \log p(\theta) + \text{const.} \; ,
\end{split}
\label{eq:loss}
\end{align}
where we first use Bayes' theorem, then ignore all terms irrelevant
for the minimization, and finally apply a simple coordinate transformation
for the bijective mapping. The last term is a simple weight
regularization, while the first two terms are called the maximum
likelihood loss. Since we impose the latent distribution of the random
variable $p(g(x_p, x_d))$ to produce a normal distribution centered
around zero and with width one, the first term becomes
\begin{align}
  \log p(g(x_p,x_d)) = -\frac{||g(x_p, x_d))||_2^2}{2} \; .
\end{align}
The final network setup after tuning of the hyper-paramaeters are liste In Tab.~\ref{tab:cinn}. We verified that the network performance is stable under small changes of these parameters.

%------------------------------------------------------------
\begin{figure}[t]
\includegraphics[page=9, width=0.505\textwidth]{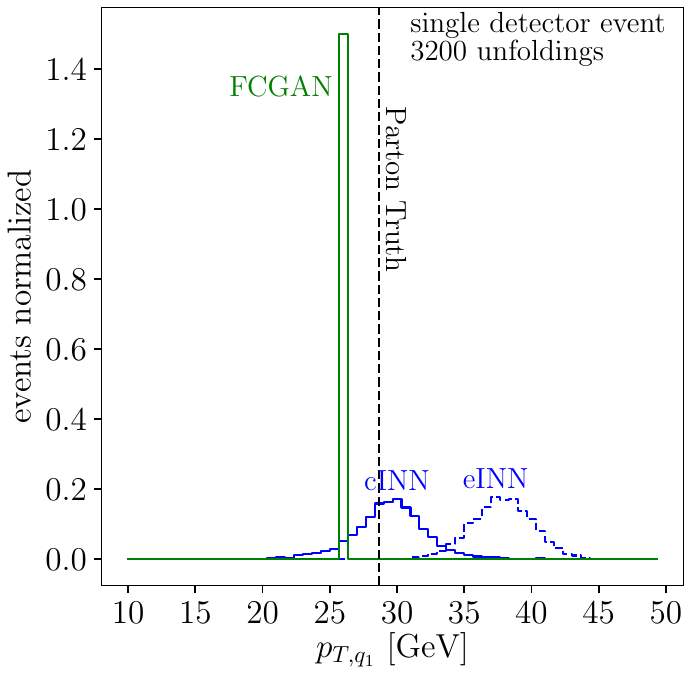}
\includegraphics[page=20, width=0.495\textwidth]{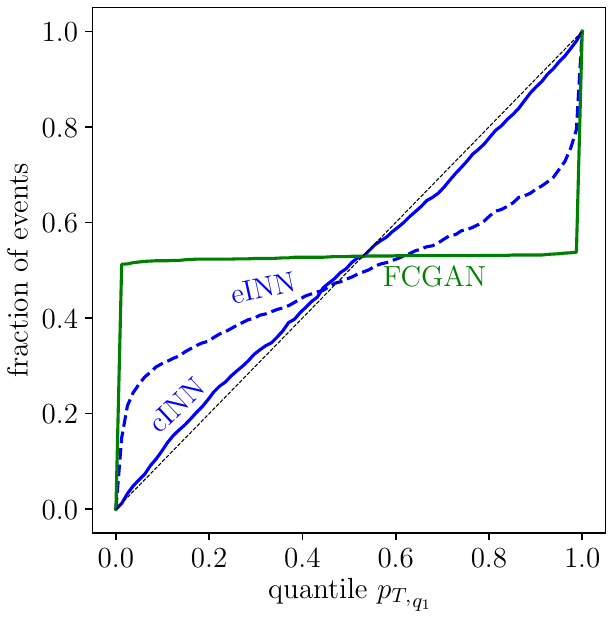}
\caption{Left: illustration of the statistical interpretation of
  unfolded events for one event. Right: calibration curves for
  $p_{T,q_1}$ extracted from the FCGAN and the noise-extended eINN, as
  shown in Fig.~\ref{fig:quantile1}, and the cINN.}
\label{fig:quantile2}
\end{figure}
%------------------------------------------------------------

In the left panels of Fig.~\ref{fig:2j} we show the unfolding
performance of the cINN, trained and tested on the same exclusive
2-jet events as the simpler INNs in Fig.~\ref{fig:UnfoldCurve}.
Unlike the naive and the noise-extended INNs we cannot evaluate the
cINN in both direction, detector simulation and unfolding, so we focus
on the detector unfolding. The agreement between parton-level truth
and the INN-unfolded distribution is around 10\% for the bulk of the
$p_T$ distributions, with the usual larger relative deviations in the
tails. An interesting feature is still the cut $p_{T,j} > 20$~GeV at
the detector level, because it leads to a slight shift in the peak of
the $p_{T,j_2}$ distribution.  Finally, the reconstructed invariant
$W$-mass and the physical $W$-width agree extremely well with the Monte
Carlo truth owing to the MMD loss.\medskip

As in Fig.~\ref{fig:quantile1} we can interpret the unfolding output
for a given detector-level event statistically. First, in the left
panel of Fig.~\ref{fig:quantile2} we show a single event and how the
FCGAN, INN, and cINN output is distributed in parton level phase
space\footnote{Throughout this paper we only compare to the FCGAN
  analysis~\cite{fcgan}, which we fully control. For standard
  unfolding methods used by ATLAS and CMS and for the new Omnifold
  method~\cite{Andreassen:2019cjw} we refrain from comments which
  would need to be based on an in-depth comparison.}. The separation
between truth and sampled distributions does not have any
significance, but we see that the cINN inherits the beneficial
features of the noise-extended eINN.  In the right panel of
Fig.~\ref{fig:quantile2} we again reconstruct the individual
probability distribution from the unfolding numerically. We then
determine the position of the parton-level truth in its respective
probability distribution for the INN and the cINN. We expect a given
percentage of the 1500 events to fall into the correct quantile of its
respective probability distribution. The corresponding calibration
curve for the cINN is added to the right panel of
Fig.~\ref{fig:quantile2}, indicating that without additional
calibration the output of the cINN unfolding can be interpreted as a
probability distribution in parton-level phase space for a single
detector-level event, as always assuming an unfolding model. Instead
of the transverse momentum of the harder parton-level quark we could
use any other kinematic distribution at parton level. This marks the
final step for a statistically interpretable unfolding.

%%%%%%%%%%%%%%%%%%%%%%%%%%%%%%%%%%%%%%%%%%%%%%%%%%%%%%%%%
\section{Unfolding with jet radiation}
\label{sec:jets}

In the previous chapter we use a simplified data set to explore
different possibilities to unfold detector level information with
invertible networks. We limit the data to events with exactly two
jets, by switching off initial state radiation (ISR). This guarantees
that the two jets come from the $W$-decay, so the network does not
have to learn this feature. In a realistic QCD environment we do not
have that information, because additional QCD jets will be radiated
off the initial and final state partons. In this section we
demonstrate how we can unfold a sample of events including ISR and
hence with a variable number of jets. We know that with very few
exceptions~\cite{Plehn:2001nj,Buckley:2014fqa} the radiation of QCD
jets does not help us understand the nature of the hard process. In
such cases, we would like to interpret a measurement with an
appropriately defined hard process, leading to the question if an
unfolding network can invert detector effects and QCD jet
radiation. Technically, this means inverting jet radiation and
kinematic modifications to the hard process as, in our case, done by
\textsc{Pythia}.

We emphasize that this approach requires us to define a specific hard
process with any number of external jets and other features. We can
illustrate this choice for two examples. First, a di-tau resonance
search typically probes the hard process $pp \to \mu^+ \mu^-+X$, where
$X$ denotes any number of additional, analysis-irrelevant jets. We
invert the corresponding measurements to the partonic process $pp \to
\mu^+ \mu^-$. A similar mono-jet analysis instead probes the process
$pp \to Z' j (j) +X$, where $Z'$ is a dark matter mediator decaying to
two invisible dark matter candidate. Depending on the analysis, the
relevant process to invert is $pp \to Z' j$ or $pp \to Z' jj$, where a
reported missing transverse momentum recoils against one or two hard
jets. Because our inversion network in trained on Monte Carlo data, we
automatically define the appropriate hard process when generating the
training data. This covers any combination of signal and background
matrix elements contributing to such a hard process, even non-SM
processes to quantify a remaining model dependence. A final caveat ---
in the hard process we do not include subjet aspects at this stage. As
long as subjet information is used for tagging purposes it factorizes
from the hard process information and can easily be included in terms
of efficiencies. A problem would arise in unfolding or inverting
analyses relying on different hard processes, like a fat mono-jet
analysis, where the above choice of recoil jets is left to a sub-jet
algorithm.

%------------------------------------------------------------
\begin{table}[b!]
\centering
\begin{small} \begin{tabular}{l|c c}
\toprule
Parameter & cINN no ISR& cINN ISR incl.  \\
\midrule
Blocks & 24 & 24 \\
Layers per block & 2 & 3 \\
Units per layer & 256 & 256 \\
Condition/encoder layers & 2 & 8 \\
Units per condition/encoder layer & 1024 & 1024 \\
Condition/encoder output dimension & 256 & 256 \\
Trainable weights & $\sim$ 2 M & $\sim$ 10 M \\
Encoder pre training epochs & - & 300\\
Epochs & 1000 & 900 \\
Learning rate & $8 \cdot 10^{-4}$ & $8 \cdot 10^{-4}$\\
Batch size & 512 & 512 \\
Training/testing events & 290k / 30k & 620k / 160k\\
Kernel widths & $\sim 2, 8, 25, 67$ & $\sim 2, 8, 25, 67$\\
$D_p$ & 12 & 12 \\
$D_d$ & 16 & 25 \\
$\lambda_\text{MMD}$ & 0.5 & 0.04 \\
$\lambda_\text{MMD}$ increase & - & 1.6 / 100 epochs\\
\bottomrule
\end{tabular} \end{small}
\caption{cINN setup and hyper-parameters, as implemented in
  \pytorch(v1.2.0)~\cite{pytorch}.}
\label{tab:cinn}
\end{table}
%------------------------------------------------------------

%%%%%%%%%%%%%%%%%%%%%%%%%%%%%%%%%%%%%%%%%%%%%%%%%%%%%%%%%
\subsection{Individual $n$-jet samples}
\label{sec:jets_indiv}

%------------------------------------------------------------
\begin{figure}[t]
\includegraphics[page = 9, width=0.5\textwidth]{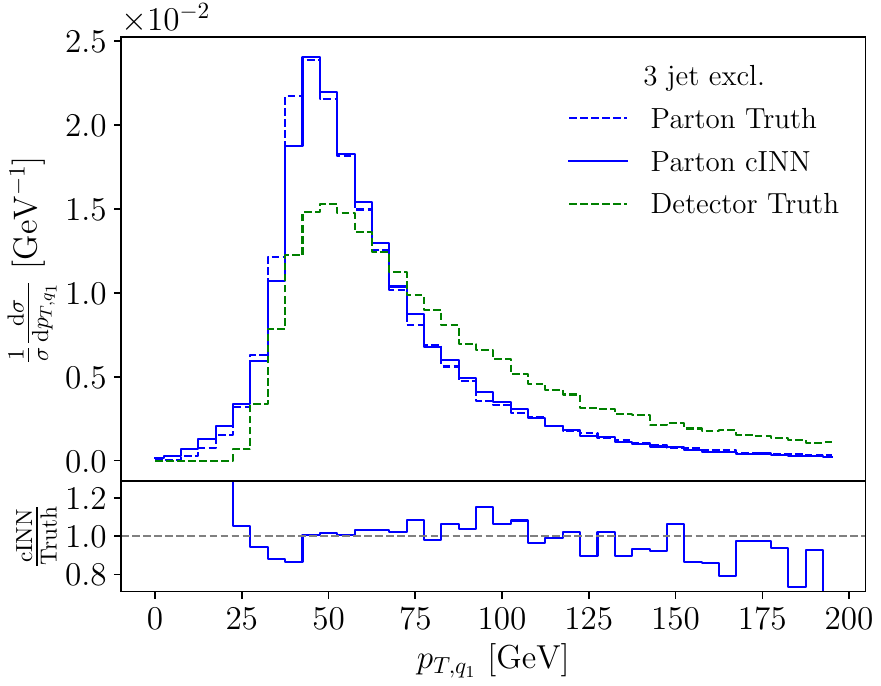}
\includegraphics[page = 9, width=0.5\textwidth]{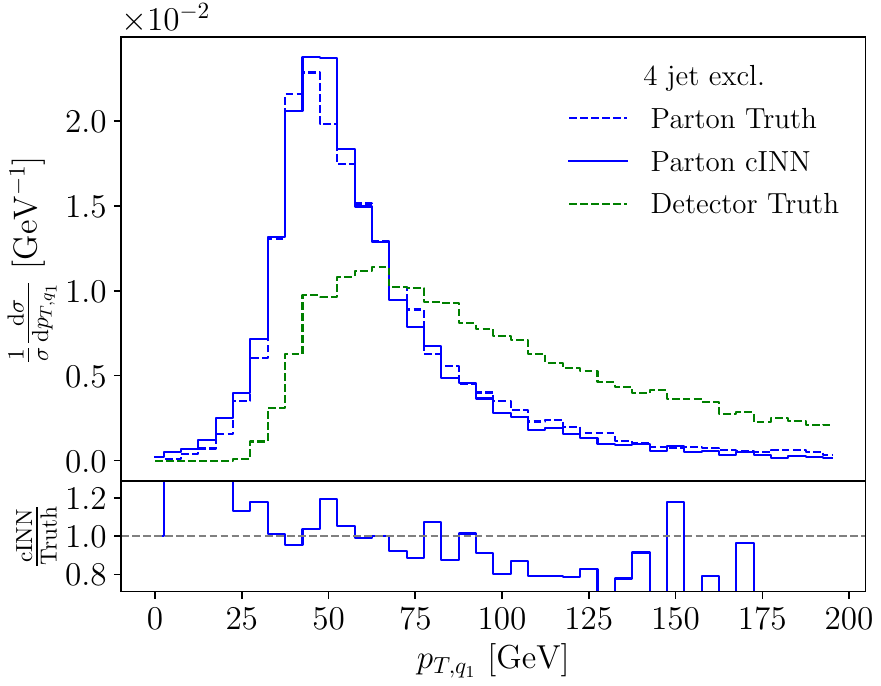} \\
\includegraphics[page =10, width=0.5\textwidth]{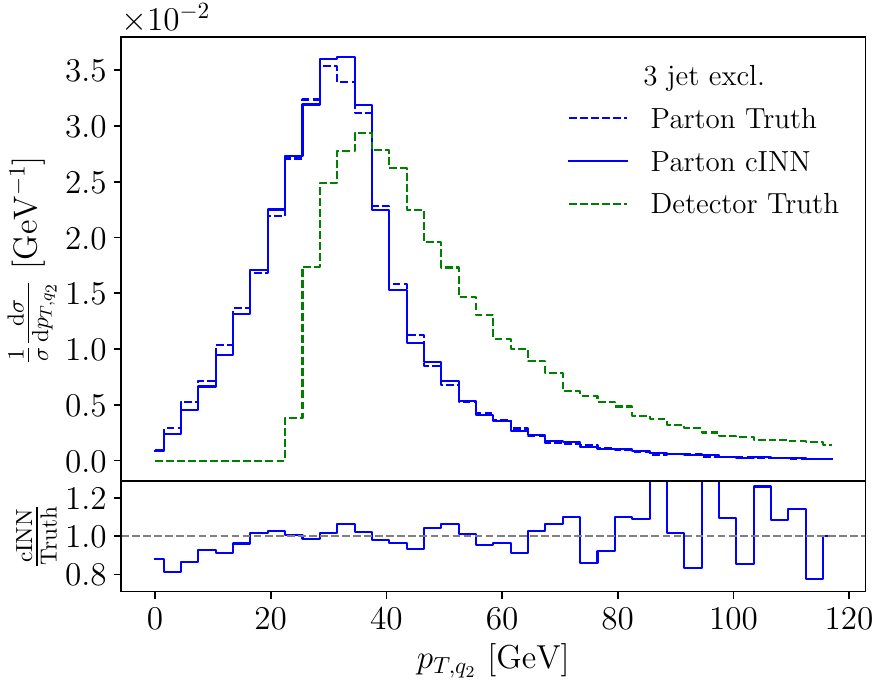} 
\includegraphics[page =10, width=0.5\textwidth]{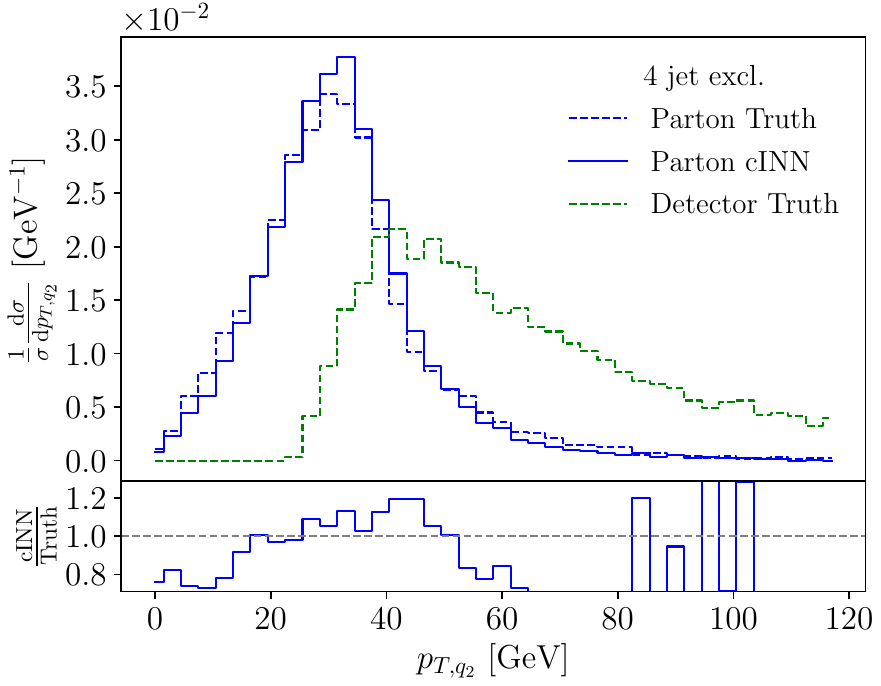} \\
\includegraphics[page =19, width=0.5\textwidth]{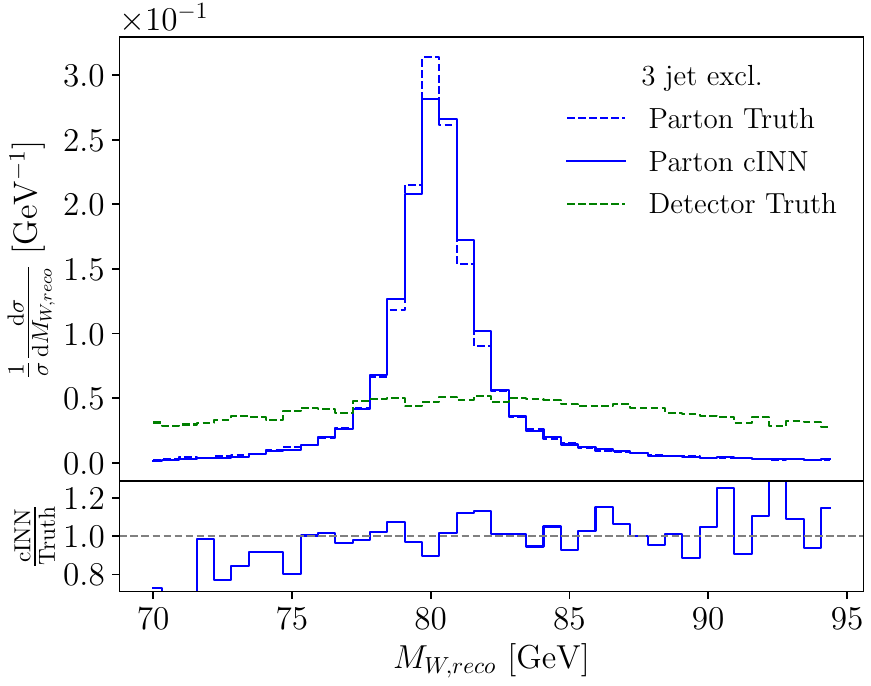}
\includegraphics[page =19, width=0.5\textwidth]{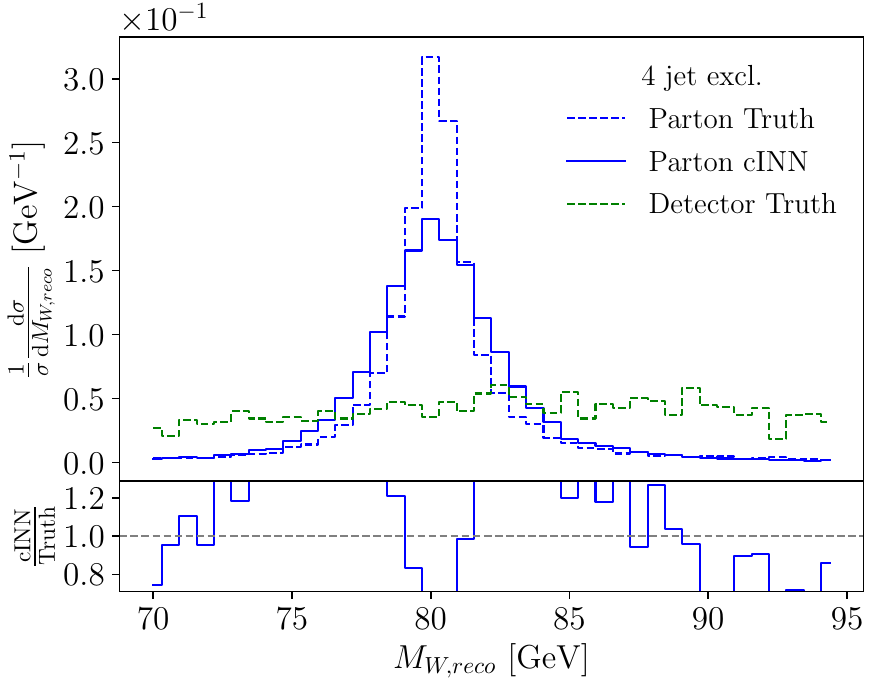}
\caption{cINNed $p_{T,q}$ and $m_{W,\text{reco}}$ distributions.  Training and
  testing events include exactly three (left) and four (right) jets
  from the data set including ISR.}
\label{fig:34j}
\end{figure}
%------------------------------------------------------------

In Sec.~\ref{sec:inn_cond} we have shown that our cINN can unfold
detector effects for $ZW$-production at the LHC.  The crucial new
feature of the cINN is that it provides probability distribution in
parton-level phase space for a given detector-level event. The actual
unfolding results are illustrated in Fig.~\ref{fig:2j}, focusing on
the two critical distribution known from the corresponding FCGAN
analysis~\cite{fcgan}.  The event sample used throughout
Sec.~\ref{sec:inn} includes exactly two partons from a $W$-decay with
minimal phase space cuts on the corresponding jets. Strictly speaking,
these phase space cuts are not necessary in this simulation. The
correct definition of a process described by perturbative QCD includes
a free number of additional jets,
\begin{align}
pp
\to ZW^\pm + \text{jets}
\to (\ell^- \ell^+) \; (j j ) + \text{jets} \; ,
\label{eq:proc_jets}
\end{align}
For the additional jets we need to include for instance a $p_T$ cut to
regularize the soft and collinear divergences at fixed-order
perturbation theory. The proper way of generating events is therefore
to allow for any number of additional jets and then cut on the number
of hard jets.  Since ISR can lead to jets with larger $p_T$ than the
$W$-decay jets, an assignment of the hardest jets to hard partons does
not work. We simply sort jets and partons by their respective $p_T$
and let the network work out their relations.  We limit the number of
jets to four because larger jet number appear very rarely and would
not give us enough training events.

Combining all jet multiplicities we use 780k events, out of which 530k
include exactly two jets, 190k events include three jets and 60k have
four or more jets. We split the data into 80\% training data and 20\%
test data to produce the shown plots.  For the network input we
zero-pad the event-vector for events with less than four jets and add
the number of jets as additional information. The training samples are
then split by exclusive jet multiplicity, such that the cINN
reconstructs the 2-quark parton-level kinematics from two, three, and
four jets at the detector level.

As before, we can start with the sample including exactly two
jets. The difference to the sample used before is that now one of the
$W$-decay jets might not pass the jet $p_T$ condition in Eq.\eqref{eq:jetcond}, so it
will be replaced by an ISR jet in the 2-jet sample. Going back to
Fig.~\ref{fig:2j} we see in the right panel how these events are
slightly different from the sample with only decay jet. The main
difference is in $p_{T,q_2}$, where the QCD radiation produces
significantly more soft jets. Still, the network learns these
features, and the unfolding for the sample without ISR and the 2-jet
exclusive sample has a similar quality. In Fig.~\ref{fig:34j} we see
the same distributions for the exclusive 3-jet and 4-jet samples. In
this case we omit the secondary panels because they are dominated by
the statistical uncertainties of the training sample. For these samples the network
has to extract the parton-level kinematics with two jets only from up
to four jets in the final state. In many cases this corresponds to
just ignoring the two softest jets and mapping the two hardest jets on
the two $W$-decay quarks, but from the $p_{T,q_2}$ distributions in
Fig.~\ref{fig:2j} we know that this is not always the correct
solution. Especially in the critical $m_{jj}$ peak reconstruction we
see that the network feels the challenge, even though the other
unfolded distributions look fine.

%%%%%%%%%%%%%%%%%%%%%%%%%%%%%%%%%%%%%%%%%%%%%%%%%%%%%%%%%
\subsection{Combined $n$-jet sample}
\label{sec:jets_all}

%------------------------------------------------------------
\begin{figure}[t]
\includegraphics[page = 9, width=0.5\textwidth]{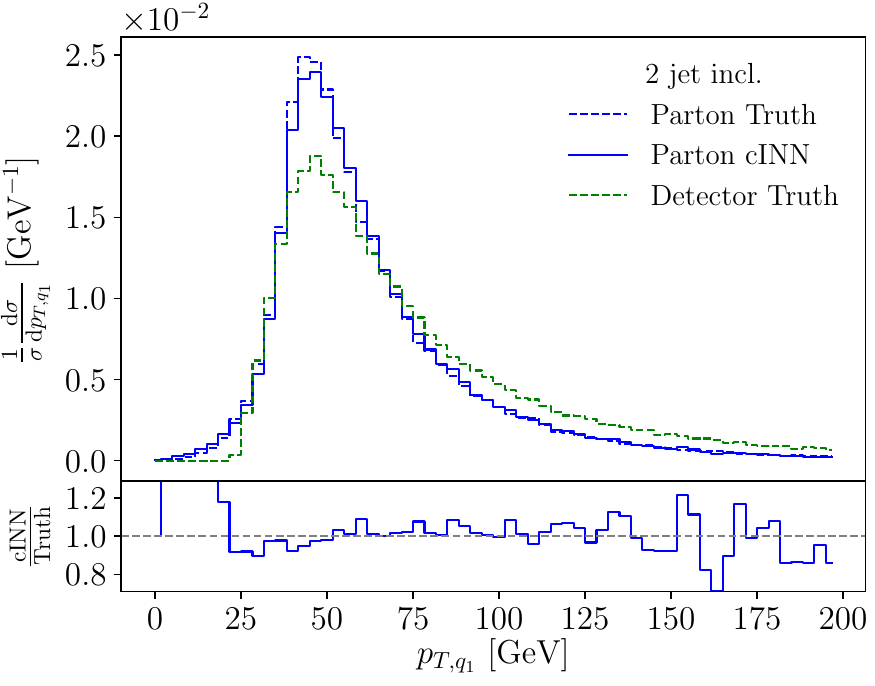}
\includegraphics[page =10, width=0.5\textwidth]{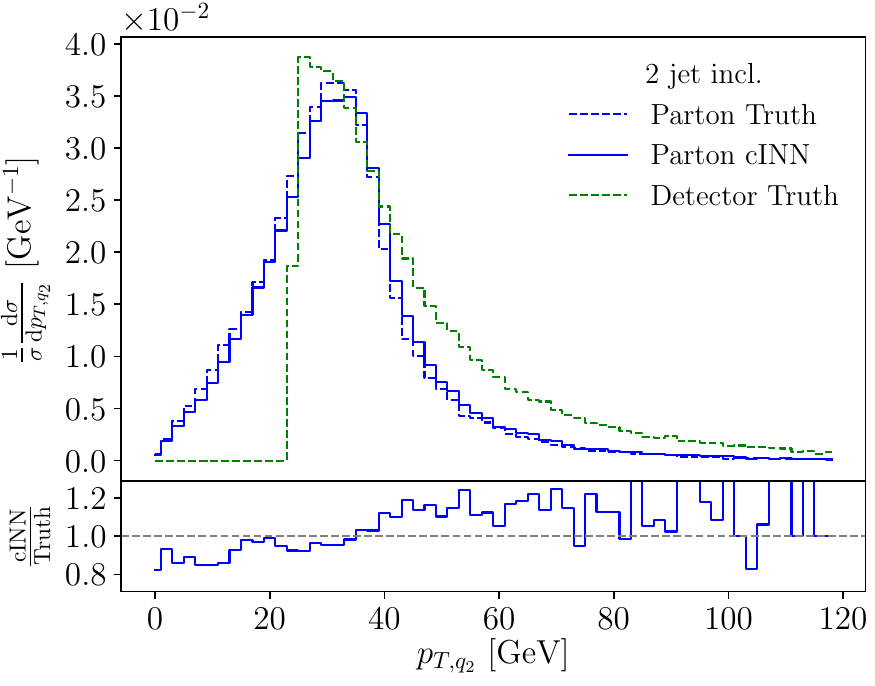} \\
\includegraphics[page =13, width=0.5\textwidth]{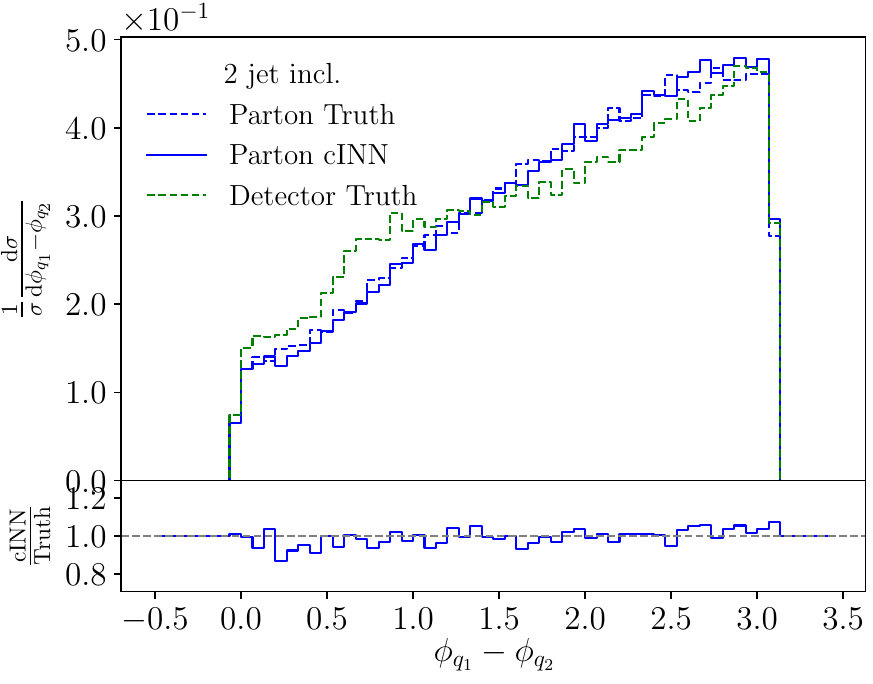}
\includegraphics[page =19, width=0.5\textwidth]{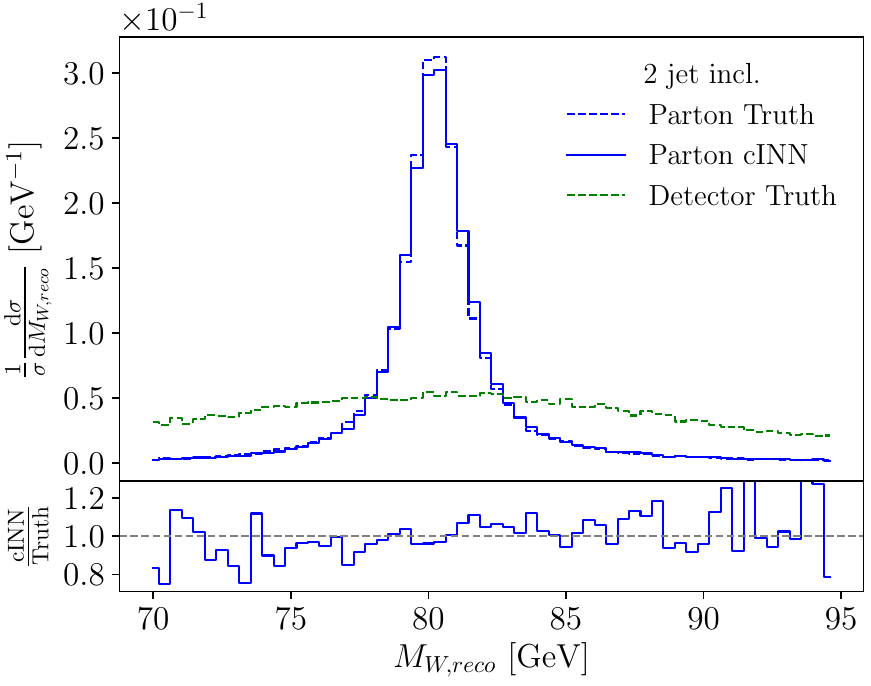}
\caption{cINNed example distributions. Training and testing events
  include two to four jets, combining the samples from
  Fig.~\ref{fig:2j} and Fig.~\ref{fig:34j} in one network. At the
  parton level there exist only two $W$-decay quarks.}
\label{fig:allj}
\end{figure}
%------------------------------------------------------------

The obvious final question is if our INN can also reconstruct the hard
scattering process with its two $W$-decay quarks from a sample with a variable number of
jets. Instead of separate samples as in Sec.~\ref{sec:jets_indiv} we
now interpret the process in Eq.\eqref{eq:proc_jets} as
jet-inclusive. This means that the hard process includes only the two
$W$-decay jets, and all additional jets are understood as jet radiation,
described either by resummed ISR or by fixed-order QCD corrections.
The training sample consists of the combination of the right panels in
Fig.~\ref{fig:2j} and the two panels in Fig.~\ref{fig:34j}. This means
that the network has to deal with the different number of jets in the
final state and how they can be related to the two hard jets of the
partonic $ZW \to \ell \ell jj$ process. The number of jets in the
final state is not given by individual hard partons, but by the jet
algorithm and its $R$-separation.

In Fig.~\ref{fig:allj} we show a set of unfolded distributions. First,
we see that the $p_{T,j}$ thresholds at the detector level are
corrected to allow for $p_{T,q}$ values to zero.  Next, we see that
the comparably flat azimuthal angle difference at the parton level is
reproduced to better than 10\% over the entire range. Finally, the
$m_{jj}$ distribution with its MMD loss re-generates the $W$-mass peak
at the parton level almost perfectly. The precision of this unfolding
is not any worse than it is for the case where the number of hard
partons and jets have to match and we only unfold the detector
effects.

%------------------------------------------------------------
\begin{figure}[t]
\includegraphics[page = 5, width=0.5\textwidth]{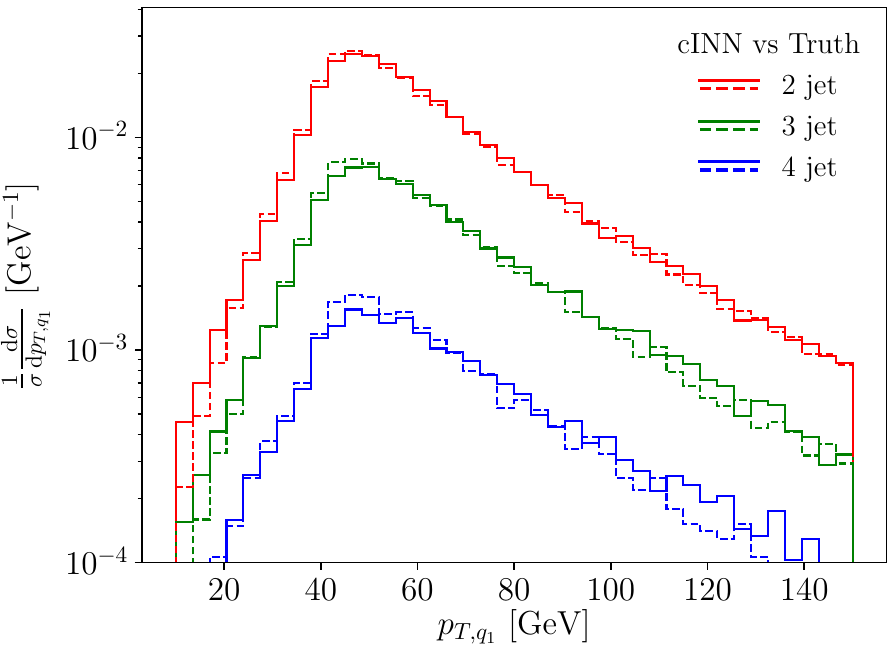}
\includegraphics[page =10, width=0.5\textwidth]{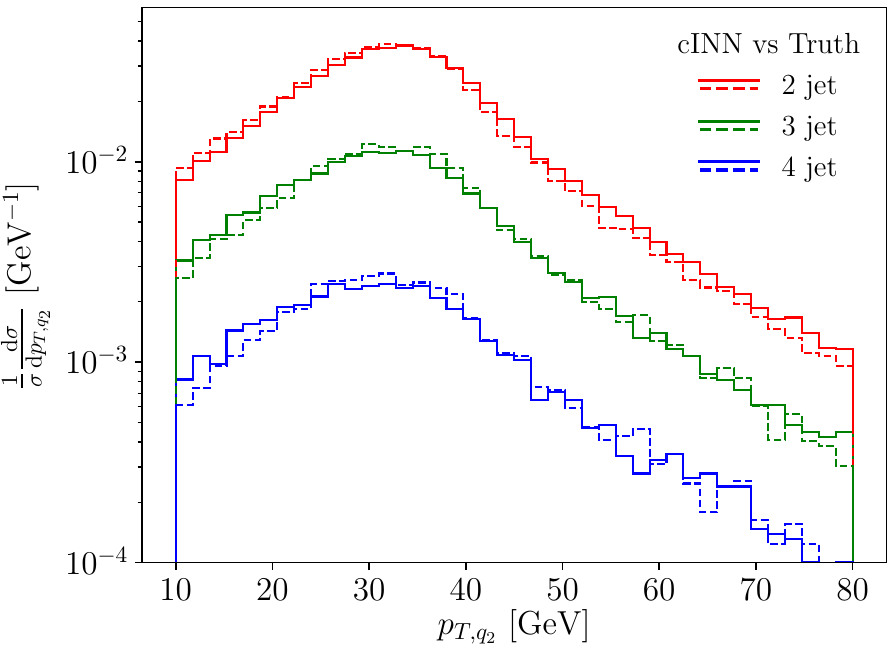} \\
\includegraphics[page =12, width=0.5\textwidth]{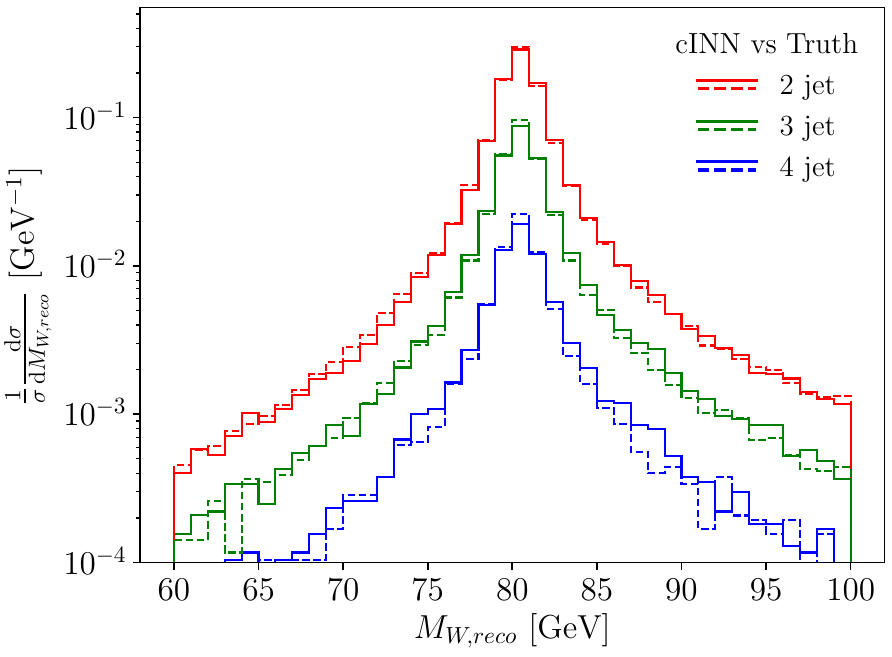}
\includegraphics[page =13, width=0.5\textwidth]{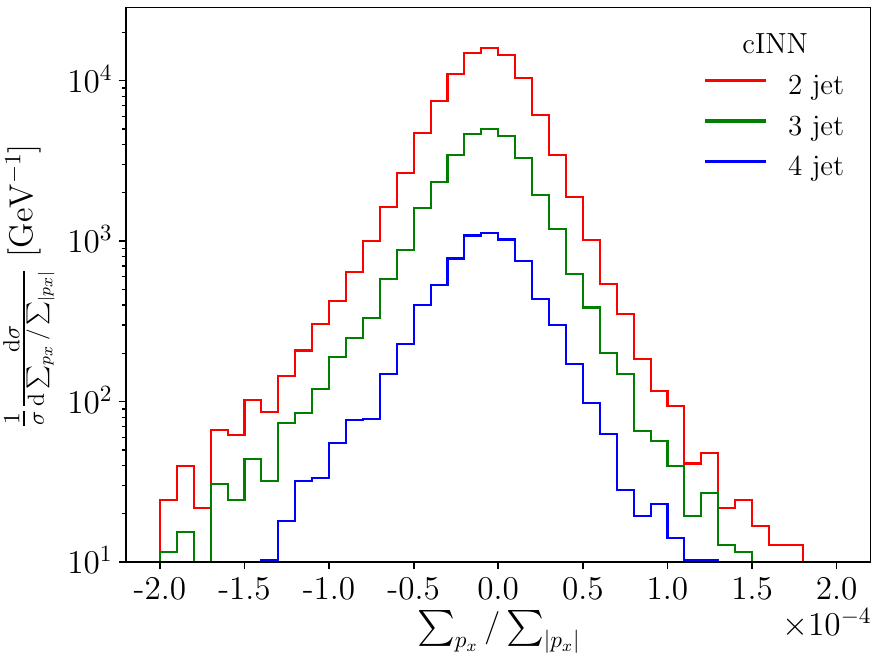}
\caption{cINNed example distributions. Training and testing events
  include two to four events, combining the samples from
  Fig.~\ref{fig:2j} and Fig.~\ref{fig:34j} in one network. The
  parton-level events are stacked by number of jets at detector
  level.}
\label{fig:stacked}
\end{figure}
%------------------------------------------------------------

In Fig.~\ref{fig:stacked} we split the unfolded distributions in
Fig.~\ref{fig:allj} by the number of 2, 3, and 4 jets in the
detector-level events. In the first two panels we see that the
transverse momentum spectra of the hard partons are essentially
independent of the QCD jet radiation. In the language of higher-order
calculations this means that we can describe extra jet radiation with
a constant $K$-factor, if necessary with the appropriate phase space
mapping. Also the reconstruction of the $W$-mass is not affected by
the extra jets, confirming that the neural network correctly
identifies the $W$-decay jets and separates them from the ISR
jets. Finally, we test the transverse momentum conservation at the
unfolded parton level. Independent of the number of jets in the final
state the energy and momentum for the pre-defined hard process is
conserved at the $10^{-4}$ level. The kinematic modifications from the
ISR simulation are unfolded correctly, so we can compute the matrix
element for the hard process and use it for instance for inference.\medskip

%%%%%%%%%%%%%%%%%%%%%%%%%%%%%%%%%%%%%%%%%%%%%%%%%%%%%%%%%%%%%%%%%%%%%%
\section{Outlook}

We have shown how an invertible network (INN) and in particular a
conditional INN can be used to unfold detector effects for the simple
example process of $ZW \to \ell \ell jj$ production at the LHC. The
cINN is not only able to unfold the process over the entire phase
space, it also gives correctly calibrated posterior probability
distributions over parton-level phase space for given detector-level
events. This feature is new even for neural network unfolding.

Next, we have extended the unfolding to a variable number of jets in
the final state. This situation will automatically appear whenever we
include higher-order corrections in perturbative QCD for a given hard
process. The hard process at parton level is defined at the training
level. We find that the cINN also unfolds QCD jet radiation in the
sense that it identifies the ISR jets and corrects the kinematics of
the hard process to ensure energy-momentum conservation in the hard
scattering.

In combination, these features should enable analysis techniques like
the matrix element method and efficient ways to communicate analysis
results including multi-dimensional kinematic distributions. While the
$ZW$ production process used in this analysis, we expect these results
to carry over to more complex processes with intermediate
particles~\cite{gan_phasespace} and the impact of a SM-training
hypothesis should be under control~\cite{fcgan}, the next step will be
to test this new framework in a realistic LHC example with proper
precision predictions and a focus on uncertainties. As for any
analysis method suitable for the coming LHC runs, the challenge will
be to control the full uncertainty budget at the per-cent
level.\footnote{We are very happy to share our code upon request, if
  colleagues are interested in tackling any such open questions.}

%%%%%%%%%%%%%%%%%%%%%%%%%%%%%%%%%%%%%%%%%%%%%%%%%%%%%%%%%%%%%%%%%%%%%%
\begin{center} \textbf{Acknowledgments} \end{center}

We would like to thank Ben Nachman for great discussions and
Hans-Christian Schultz-Coulon for the experimental encouragement.  RW
and MB acknowledge support by the IMPRS-PTFS.  The research of AB, MB,
and TP is supported by the Deutsche Forschungsgemeinschaft (DFG,
German Research Foundation) under grant 396021762 -- TRR~257
\textsl{Particle Physics Phenomenology after the Higgs Discovery}. GK
acknowledges support by the DFG under Germany’s Excellence Strategy –
EXC 2121 \textsl{Quantum Universe – 390833306}.

\end{fmffile}
%%%%%%%%%%%%%%%%%%%%%%%%%%%%%%%%%%%%%%%%%%%%%%%%%%%%%%%%%%%%%%%%%%%%%%%%

\bibliography{literature}

\end{document}

%% file: incl_declare.tex
\setitemize{itemsep=2pt,topsep=2pt,parsep=0pt,partopsep=0pt,leftmargin=*}
\setenumerate{itemsep=0pt,topsep=2pt,parsep=0pt,partopsep=0pt,labelindent=3pt,leftmargin=*}

\definecolor{Gcolor}{HTML}{3b528b}
\definecolor{Dcolor}{HTML}{e41a1c}

\tikzstyle{generator} = [rectangle, rounded corners, minimum width=3cm, minimum height=1cm,text centered, draw=Gcolor]
\tikzstyle{discriminator} = [rectangle, rounded corners, minimum width=3cm, minimum height=1cm,text centered, draw=Dcolor]
\tikzstyle{io} = [circle, trapezium left angle=70, trapezium right angle=110, minimum width=1cm, minimum height=1cm, text centered, draw=black]

\tikzstyle{process} = [rectangle, minimum width=1cm, minimum height=1cm, text centered, draw=black]
\tikzstyle{decision} = [rectangle, minimum width=1cm, minimum height=1cm, text centered, draw=black]

\tikzstyle{arrow} = [thick,->,>=stealth]
\usepackage{xcolor}

\newcommand\one{\leavevmode\hbox{\small1\normalsize\kern-.33em1}}

\newcommand{\gev}{\text{GeV}}

\def\slashchar#1{\setbox0=\hbox{$#1$}           % set a box for #1
   \dimen0=\wd0                                 % and get its size
   \setbox1=\hbox{/} \dimen1=\wd1               % get size of /
   \ifdim\dimen0>\dimen1                        % #1 is bigger
      \rlap{\hbox to \dimen0{\hfil/\hfil}}      % so center / in box
      #1                                        % and print #1
   \else                                        % / is bigger
      \rlap{\hbox to \dimen1{\hfil$#1$\hfil}}   % so center #1
      /                                         % and print /
   \fi}

\newcommand{\ie}{\textsl{i.e.}\;}

\renewcommand{\d}{{\text{d}}}

\newcommand{\madgraph}{\textsc{Madgraph}5\xspace}
\newcommand{\pythia}{\textsc{Pythia8}\xspace}
\newcommand{\fastjet}{\textsc{FastJet}\xspace}
\newcommand{\delphes}{\textsc{Delphes}\xspace}

\newcommand{\pytorch}{\textsc{PyTorch}\xspace}

%% file: incl_feynman.tex
\begin{fmfgraph*}(140,70)
\fmfset{arrow_len}{2mm}
\fmfset{curly_len}{3mm}
\fmfset{wiggly_len}{3mm}
\fmfstraight
\fmfleft{i1,i2}
\fmfright{o1,o2,o3,o4}
\fmf{fermion,tension=1.0,width=0.6}{i1,v1,i2}
\fmf{photon,tension=2.5,width=0.6}{v1,v2}	
\fmf{photon,tension=1.0,label=$W$,lab.side=right,width=0.6}{v2,d1}		
\fmf{photon,tension=1.0,label=$Z$,lab.side=left,width=0.6}{v2,d2}
\fmf{fermion,width=0.6}{o3,d2,o4}
\fmf{fermion,width=0.6}{o2,d1,o1}
\fmflabel{$j$}{o1}
\fmflabel{$j$}{o2}
\fmflabel{$\ell^+$}{o3}
\fmflabel{$\ell^-$}{o4}
\end{fmfgraph*}

%% file: incl_network_inn.tex
%\definecolor{Gcolor}{HTML}{3b528b}
%\definecolor{Dcolor}{HTML}{e41a1c}

\definecolor{Gcolor}{HTML}{2c7fb8}
\definecolor{Dcolor}{HTML}{f03b20}

\tikzstyle{INN} = [thick, rectangle, rounded corners, minimum width=2.5cm, minimum height=3.5cm,text centered, draw=Gcolor]
\tikzstyle{preprocessor} = [thick, rectangle, rounded corners, minimum width=1.5cm, minimum height=1cm,text centered, draw=Dcolor]
\tikzstyle{mmd} = [thick, rectangle, rounded corners, minimum width=1.5cm, minimum height=1cm,text centered, draw=black]
\tikzstyle{io} = [thick,circle, minimum width=1.2cm, minimum height=1.2cm, text centered, draw=black]

\tikzstyle{cond} = [thick, rectangle, dotted, rounded corners, minimum width=10.0cm, minimum height=2cm,text centered, draw=gray!50!black]

\tikzstyle{iodotted} = [thick, circle, minimum width=1.2cm, minimum height=1cm, text centered, draw=black, dotted]

\tikzstyle{process} = [thick, rectangle, minimum width=1cm, minimum height=1cm, text centered, draw=black]

\tikzstyle{xG} = [thick,rectangle, minimum width=2.2cm, minimum height=3cm, text depth= 2.2cm, draw=black]
\tikzstyle{s0} = [thick,rectangle, minimum width=2cm, minimum height=3cm, text centered]
\tikzstyle{s1} = [thick, dotted, rectangle, minimum width=1.6cm, minimum height=1.1cm, text centered, draw=black]

\tikzstyle{decision} = [thick,rectangle, minimum width=1cm, minimum height=1cm, text centered, draw=black]

\tikzstyle{dots} = [circle, minimum size=2pt, inner sep=0pt,outer sep=0pt, draw=Dcolor, fill = Dcolor]

\tikzstyle{arrow} = [thick,->,>=stealth]

\begin{tikzpicture}[node distance=2cm]

\node (INN) [INN] {INN};

\node (xG) [io, left of = INN, xshift=-0.8cm, yshift=1cm] {$\{\tilde x_{p}, \tilde r_{p}\}$};
\node (rG) [io, right of = INN, xshift=0.8cm, yshift=-1cm] {$\{\tilde x_{d}, \tilde r_{d}\}$};

\node (xp) [io, left of = INN, xshift=-0.8cm, yshift=-1cm] {$\{x_p, r_p\}$};
\node (parton) [process, below of=xp, xshift=0cm, yshift=0cm] {parton};
\node (mmd) [mmd, left of=xp, xshift=0.0cm, yshift=1cm] {$L_\text{MMD, MSE}$};

\node (random) [io, right of=INN, xshift=0.8cm, yshift=1cm] {$\{ x_d, r_d \}$};
\node (detector) [process, above of=random, xshift=0cm, yshift=0cm] {detector};
\node (gauss) [mmd, right of=rG, xshift=0.0cm, yshift=1cm] {$L_\text{MMD, MSE}$};

%\node (cond) [cond, above of = xG, xshift=1.5cm, yshift=-0.1cm] {};
%\node (condi) [above of = xG, xshift=1.9cm, yshift=0.5cm, color=gray!50!black] {Condition};

%\node (preprocessor) [preprocessor, above of=INN, xshift=0cm, yshift=0.5cm] {Subnet};
%\node (xd) [io, left of = preprocessor, xshift=-0.5cm, yshift=0cm] {$\{x_d\}$};
%\node (detector) [process, left of=xd, xshift=-0.2cm, yshift=0cm] {detector};

\coordinate[ right of = rG, xshift=0cm, yshift=0cm] (Gin1);
\coordinate[ right of = random, xshift=0cm, yshift=0cm] (Gin2);

\coordinate[ left of = xG, xshift=0cm, yshift=0cm] (MMDin1);
\coordinate[ left of = xp, xshift=0cm, yshift=0cm] (MMDin2);

%\draw [arrow, color=black] ([yshift=0em]random.west) -- ([yshift=1.5em]INN.east);
%\draw [arrow, color=black] ([yshift=-1.5em]INN.east) -- (rG.west);
%\draw [arrow, color=black] ([yshift=1.5em]INN.west) -- (xG.east);
%\draw [arrow, color=black] ([yshift=0em]xp.east) -- ([yshift=-1.5em]INN.west);

\draw [arrow, color=black] ([yshift=0em]parton.north) -- ([yshift=0em]xp.south);
\draw [arrow, color=black] ([yshift=0em]detector.south) -- ([yshift=0em]random.north);

\draw [thick, color=Gcolor] ([yshift=0em]xp.west) -- ([yshift=0em]MMDin2);
\draw [arrow, color=Gcolor] ([yshift=0em]MMDin2) -- ([yshift=0em]mmd.south);
\draw [thick, color=Gcolor] ([yshift=0em]xG.west) -- ([yshift=0em]MMDin1);
\draw [arrow, color=Gcolor] ([yshift=0em]MMDin1) -- ([yshift=0em]mmd.north);

\draw [thick, color=Gcolor] ([yshift=0em]random.east) -- ([yshift=0em]Gin2);
\draw [arrow, color=Gcolor] ([yshift=0em]Gin2) -- ([yshift=0em]gauss.north);
\draw [thick, color=Gcolor] ([yshift=0em]rG.east) -- ([yshift=0em]Gin1);
\draw [arrow, color=Gcolor] ([yshift=0em]Gin1) -- ([yshift=0em]gauss.south);

%\draw [arrow, color=black] ([yshift=0em]detector.east) -- ([yshift=0em]xd.west);
%\draw [arrow, color=black] ([yshift=0em]xd.east) -- ([yshift=0em]preprocessor.west);
%\draw [arrow, color=black] ([yshift=0em]preprocessor.south) -- ([yshift=0em]INN.north);

%\draw[arrow, thick, color=Gcolor] (random.west) -- (xG.east);
\draw[arrow, color=Gcolor] (random.west) --  node[scale=0.8, sloped, anchor=center, above, color=Gcolor]{$\bar{g}(x_d, r_d)$} ([yshift=0cm]xG.east);
%\draw[arrow, thick, color=Gcolor] ([yshift=1cm]INN.west) -- (xG.east);
%\draw[arrow, thick, color=Gcolor] (xp.east) -- (rG.west);
%\draw[arrow, thick, color=Gcolor] (xp.east) -- ([yshift=-1cm]INN.west);
\draw[arrow, color=Gcolor] ([yshift=0cm]xp.east) --  node[scale=0.8, sloped, anchor=center, above, color=Gcolor]{$g(x_p, r_p)$} (rG.west);

\draw[arrow, thick, dashed, color=black] (mmd) -- (INN.west);
\draw[arrow, thick, dashed, color=black] (gauss) -- (INN.east);

\end{tikzpicture}

%% file: incl_network_cinn.tex
%\definecolor{Gcolor}{HTML}{3b528b}
%\definecolor{Dcolor}{HTML}{e41a1c}

\definecolor{Gcolor}{HTML}{2c7fb8}
\definecolor{Dcolor}{HTML}{f03b20}

\tikzstyle{cINN} = [thick, rectangle, rounded corners, minimum width=2.5cm, minimum height=3.5cm,text centered, draw=Gcolor]
\tikzstyle{preprocessor} = [thick, rectangle, rounded corners, minimum width=1.5cm, minimum height=1cm,text centered, draw=Dcolor]
\tikzstyle{mmd} = [thick, rectangle, rounded corners, minimum width=1.5cm, minimum height=1cm,text centered, draw=black]
\tikzstyle{io} = [thick,circle, minimum width=1.2cm, minimum height=1cm, text centered, draw=black]

\tikzstyle{cond} = [thick, rectangle, dotted, rounded corners, minimum width=10.0cm, minimum height=2cm,text centered, draw=gray!50!black]

\tikzstyle{iodotted} = [thick, circle, minimum width=1.2cm, minimum height=1cm, text centered, draw=black, dotted]

\tikzstyle{process} = [thick, rectangle, minimum width=1cm, minimum height=1cm, text centered, draw=black]

\tikzstyle{xG} = [thick,rectangle, minimum width=2.2cm, minimum height=3cm, text depth= 2.2cm, draw=black]
\tikzstyle{s0} = [thick,rectangle, minimum width=2cm, minimum height=3cm, text centered]
\tikzstyle{s1} = [thick, dotted, rectangle, minimum width=1.6cm, minimum height=1.1cm, text centered, draw=black]

\tikzstyle{decision} = [thick,rectangle, minimum width=1cm, minimum height=1cm, text centered, draw=black]

\tikzstyle{dots} = [circle, minimum size=2pt, inner sep=0pt,outer sep=0pt, draw=Dcolor, fill = Dcolor]

\tikzstyle{arrow} = [thick,->,>=stealth]

\begin{tikzpicture}[node distance=2cm]

\node (cINN) [cINN] {cINN};

\node (xG) [io, left of = cINN, xshift=-0.5cm, yshift=1cm] {$\{\tilde{x}_p\}$};
\node (rG) [io, right of = cINN, xshift=0.5cm, yshift=-1cm] {$\{ \tilde{r} \}$};

\node (xp) [io, left of = cINN, xshift=-0.5cm, yshift=-1cm] {$\{x_p\}$};
\node (parton) [process, below of=xp, xshift=0cm, yshift=0.25cm] {parton};
\node (mmd) [mmd, left of=xp, xshift=0.0cm, yshift=1cm] {$L_\text{MMD}$};

\node (random) [io, right of=cINN, xshift=0.5cm, yshift=1cm] {$\{ r \}$};
\node (gauss) [mmd, right of=rG, xshift=0.0cm, yshift=1cm] {$L$};

\node (cond) [cond, above of = xG, xshift=1.5cm, yshift=0.7cm] {};
\node (condi) [above of = xG, xshift=5.5cm, yshift=1.3cm, color=gray!50!black] {condition};

\node (preprocessor) [preprocessor, above of=cINN, xshift=0cm, yshift=1.5cm] {subnet};
\node (xd) [io, left of = preprocessor, xshift=-0.5cm, yshift=0cm] {$\{x_d\}$};
\node (detector) [process, left of=xd, xshift=-0.2cm, yshift=0cm] {detector};

\coordinate[ right of = rG, xshift=0cm, yshift=0cm] (Gin1);
\coordinate[ right of = random, xshift=0cm, yshift=0cm] (Gin2);

\coordinate[ left of = xG, xshift=0cm, yshift=0cm] (MMDin1);
\coordinate[ left of = xp, xshift=0cm, yshift=0cm] (MMDin2);

%\draw [arrow, color=black] ([yshift=0em]random.west) -- ([yshift=1.5em]cINN.east);
%\draw [arrow, color=black] ([yshift=-1.5em]cINN.east) -- (rG.west);
%\draw [arrow, color=black] ([yshift=1.5em]cINN.west) -- (xG.east);
%\draw [arrow, color=black] ([yshift=0em]xp.east) -- ([yshift=-1.5em]cINN.west);

\draw [arrow, color=black] ([yshift=0em]parton.north) -- ([yshift=0em]xp.south);

\draw [thick, color=Gcolor] ([yshift=0em]xp.west) -- ([yshift=0em]MMDin2);
\draw [arrow, color=Gcolor] ([yshift=0em]MMDin2) -- ([yshift=0em]mmd.south);
\draw [thick, color=Gcolor] ([yshift=0em]xG.west) -- ([yshift=0em]MMDin1);
\draw [arrow, color=Gcolor] ([yshift=0em]MMDin1) -- ([yshift=0em]mmd.north);

%\draw [thick, color=Gcolor] ([yshift=0em]random.east) -- ([yshift=0em]Gin2);
%\draw [arrow, color=Gcolor] ([yshift=0em]Gin2) -- ([yshift=0em]gauss.north);
\draw [thick, color=Gcolor] ([yshift=0em]rG.east) -- ([yshift=0em]Gin1);
\draw [arrow, color=Gcolor] ([yshift=0em]Gin1) -- ([yshift=0em]gauss.south);

\draw [arrow, color=black] ([yshift=0em]detector.east) -- ([yshift=0em]xd.west);
\draw [arrow, color=Dcolor] ([yshift=0em]xd.east) -- ([yshift=0em]preprocessor.west);
\draw [arrow, color=Dcolor] ([yshift=0em, xshift=1mm]preprocessor.south) -- node[scale=0.8, anchor=center, right, color=Dcolor]{$f(x_d)$} ([yshift=0em,xshift=1mm]cINN.north);
\draw [arrow, dashed, color=black] ([yshift=0em, xshift=-1mm]cINN.north) -- ([yshift=0em, xshift=-1mm]preprocessor.south);

\draw[arrow, thick, color=Gcolor] (random.west) -- node[scale=0.8, sloped, anchor=center, above, color=Gcolor]{$\bar{g}(r, f(x_d))$} (xG.east);
\draw[arrow, thick, color=Gcolor] (xp.east) -- node[scale=0.8, sloped, anchor=center, above, color=Gcolor]{$g(x_p, f(x_d))$} (rG.west);

\draw[arrow, thick, dashed, color=black] (mmd) -- (cINN.west);
\draw[arrow, thick, dashed, color=black] (gauss) -- (cINN.east);

\end{tikzpicture}